\documentclass[prx,aps,twocolumn,showpacs,amsmath,amssymb,superscriptaddress,floatfix]{revtex4-2}

\usepackage{graphicx}
\usepackage[export]{adjustbox}
\usepackage{dcolumn}
\usepackage{upgreek}
\usepackage{bm,dsfont}
\usepackage{amssymb,mathtools}
\usepackage{booktabs}
\usepackage{color}
\usepackage{microtype}
\usepackage[T1]{fontenc}
\usepackage{lmodern} 
\usepackage[shortlabels]{enumitem}
\usepackage{changes}
\usepackage{physics}
\usepackage{derivative}
\usepackage{pgfplots}
\usepackage{tikz}
\usepackage{soul}

\allowdisplaybreaks
\usepackage{placeins}
\usepackage[colorlinks,plainpages=false,pdfpagemode=UseNone,pdfstartview=FitBH]{hyperref}
\hypersetup{
colorlinks = true,
linkcolor = [rgb]{0.87,0.18,0.22},
citecolor = [rgb]{0.0,0.60,0.32},
urlcolor = [rgb]{0.20, 0.30, 0.54}}

\usepackage{color}
\definecolor{eggplant}{RGB}{180,33,147}
\definecolor{cayn}{rgb}{0.0,0.60,0.32}

\usepackage{sidecap}

\renewcommand{\tablename}{Tab.}
\makeatletter\renewcommand{\fnum@table}[1]{\tablename~\thetable.}\makeatother

\newcommand{\ophattext}[2]{\hat{#1}_{\text{#2} } }
\newcommand{\ophat}[2]{\hat{#1}_{#2} }

\newcommand{\vecbf}[1]{{\bf #1}}

\definecolor{citecolor}{rgb}{0.0,0.60,0.32}
\newcommand{\CCQ}{Center for Computational Quantum Physics, Flatiron Institute, 162 5th Avenue, New York, NY 10010, USA}

\begin{document}

\title{Dynamical correlation functions from complex time evolution}

\author{Xiaodong Cao}
\email{xcao@flatironinstitute.org}
\affiliation{\CCQ}

\author{Yi Lu}
\email{yilu@nju.edu.cn}
\affiliation{National Laboratory of Solid State Microstructures and Department of Physics, Nanjing University, Nanjing 210093, China}
\affiliation{Collaborative Innovation Center of Advanced Microstructures, Nanjing University, Nanjing 210093, China}

\author{E.\ Miles Stoudenmire}
\affiliation{\CCQ}

\author{Olivier Parcollet}
\affiliation{\CCQ}
\affiliation{Universit\'e Paris-Saclay, CNRS, CEA, Institut de Physique Th\'eorique, 91191, Gif-sur-Yvette, France}

\date{\today}
\begin{abstract}

We present an approach to tame the growth of entanglement during
time evolution by tensor network methods.  It combines time evolution in the complex plane with
a perturbative and controlled reconstruction of correlation functions on the real-time axis.
We benchmark our approach on the single impurity Anderson model. Compared to purely real-time evolution, the complex
time evolution significantly reduces the required bond dimension to obtain the spectral function. 
Notably, our approach yields self-energy results with high precision at low frequencies,
comparable to numerical renormalization group (NRG) results, and it successfully captures the exponentially small Kondo energy scale.
\end{abstract}
\maketitle

Spectroscopic techniques play a central role in condensed matter physics, 
as they allow one to probe various excitation spectra of the system
such as angle resolved photoemission spectra (ARPES)\cite{shen2003}, inelastic neutron scattering (INS), or
resonant inelastic x-ray scattering (RIXS)~\cite{vanden2011}.
An in-depth theoretical understanding of these experiments
strongly relies on our capability to compute the real time dynamics of the 
systems or the related frequency-dependent spectral function.
Transport properties are also directly related to such functions via the Kubo formula.

In computational quantum many-body physics, accessing time-dependent quantities in strongly correlated  systems is
often severely limited since many algorithms operate in imaginary time, such as quantum Monte Carlo.
Although Matsubara imaginary time or
frequency data provides useful information, particularly about
equilibrium properties and thermodynamics, it is challenging to use for computing
real-frequency and transport properties. 
The difficulty is due to the well-known issue of analytic continuation being numerically ill-conditioned. 
In quantum embedding methods like dynamical mean field theory (DMFT) \cite{Georges1996, Kotliar2006}, 
the lack of efficient, precise, and controlled quantum impurity solvers which can simultaneously access the low
real frequencies required for transport computations and handle large multi-orbital systems constitutes a significant bottleneck.

In recent years, progress has been made for low-dimensional systems whose ground
state can be efficiently represented by matrix product states
(MPS)~\cite{Fannes_fcs_1992,Ostlund_fcs_1995,Frank_mps_2008} and optimized
through the density matrix renormalization group (DMRG)~\cite{dmrg_white_1992, dmrg_white_1993,Ulrich_review_2011}.
Two main directions have been pursued for calculating various spectral functions.
First, computational methods have been developed which work directly in frequency space
including the Lanzcos-vector method~\cite{Hallberg_lanczos_1995,Dargel_lanczos_refined_2011,Dargel_lanczos_refined_2012},
correction vector method~\cite{White_cv_1999}, 
dynamical DMRG (DDMRG)~\cite{Jeckelmann_ddmrg_2002,Jeckelmann_ddmrg_2008,Weichselbaum_unfiying_nrg_vmps_2009,Jeckelmann_deconvolution_blind_2014,Alvarez_cv_krylov_2016,Alvarez_cv_rootN_2022},
and Chebyshev polynomials in combination with
MPS~\cite{Holzner_chebyshev_spectra_2011, Wolf_chebyshev_time_2015,
Halimeh_chebyshev_time_2015,Xiang_chebyshev_reorth_2018}.
Second, the spectral functions can also be obtained from time-dependent correlation functions,
using precise time evolution methods.
These include gate evolution based on the Suzuki-Trotter
decomposition~\cite{Vidal_cononical_form_2003,Vidal_time_evolution_2004,
White_tDMRG_2004, Daley_tebd_2004,Frank_time_finite_temp_2004,
Vidal_time_evolution_mixed_state_2004}, efficient matrix product operator
(MPO) approximations of the evolution
operator~\cite{Pollmann_MPO_time_evolution_2015}, DMRG Lanczos-vector
methods~\cite{Juan_krylov_global_2006,Dargel_krylov_global_2012,Wall_krylov_global_2012},
time-step targeting DMRG~\cite{Feiguin_tdmrg_2005, Peter_krylov_local_2004,
Rodriguez_tdmrg_2006, Garnet_tdmrg_2017}, and the time-dependent variational
principle (TDVP)~\cite{tdvp_rk_2011, tdvp_tangent_2013, tdvp_local_2016,
tdvp_tangent_uMPS_2019}, each with its strengths and
weaknesses~\cite{Paeckel_time_evolution_review_2019}. 

Both approaches encounter a similar difficulty at long times or low frequencies.
Frequency methods face limitations due to the necessity of using a broadening parameter, denoted as $\gamma$,
to regulate the computations at low frequency $\omega$.
Time evolution methods are limited at long time by the growth of the entanglement:
they require the use of a tensor network bond dimension
which grows with time~\cite{Calabrese_entangle_2005,Polkovnikov_2011, Tobias_entanglement_time_2006,
Tobias_entanglement_time_appro_2006,Bravyi_upper_bounds_unitary_evolution_2007,Frank_upper_bounds_stability_2016,Cirac_tns_local_time_evolution_2021,Kuwahara_improved_area_law_2021}, in contrast with imaginary time evolution where the bond dimension saturates.
Consequently, the long-time (and therefore low-frequency) behavior must be
inferred through extrapolation techniques, such as linear prediction
\cite{White_linear_prediction_2008, White_recursion_2021}.
As a result, MPS wavefunction algorithms are limited in their accuracy at low frequencies.
For instance, state-of-the-art quantum impurity solvers based on tensor networks cannot match
the accuracy of the numerical renormalization group (NRG) at low frequency. Since they can 
access much larger and more complicated systems, however, we can expect any significant improvement in tensor network time evolution 
to have a large impact on the field.

In this paper, we address this challenge using Hamiltonian {\it time evolution in the complex plane}, 
coupled with a straightforward but controlled perturbative technique to reconstruct the physical correlation function 
on the real time axis.
We demonstrate the feasibility of using a complex-plane contour
located at a sufficient distance from the real axis to effectively limit the
entanglement growth while remaining close enough to allow an accurate reconstruction of the
function on the real axis. 
Notably, this approach delivers a highly accurate
Fermi liquid self-energy at low frequencies, displaying excellent agreement
with NRG. Let us mention that complementary approaches to exploit the complex time idea are presented in \cite{grundner2023complex}.

This paper is organized as follows. In Sec.~\ref{sec::complex_time_evolution}
we introduce the idea of time evolution in the complex plane and show how the
spectral function is retrieved from the complex time contour.
In Sec.~\ref{sec::results_siam}, we benchmark our methods on 
the single impurity Anderson model (SIAM). Conclusions and
future directions are presented in Sec.~\ref{sec::conclusion}.


\section{Method}\label{sec::method}

In this section, we describe our method, a complex time evolution followed 
by a perturbative reconstruction of the result on the real axis.
We wish to compute the real-time correlation functions involving operators
$\ophat{O}{1}, \ophat{O}{2}$  given by
\begin{subequations}
\begin{align}\label{eq::gt_real}
   G^{>}_{\ophat{O}{1} \ophat{O}{2}}(t) &= -i \bra{\psi_g} \ophat{O}{1}(t) \ophat{O}{2} \ket{\psi_g}  \\ 
   G^{<}_{\ophat{O}{1} \ophat{O}{2}}(t) &= -\zeta i  \bra{\psi_g} \ophat{O}{2} \ophat{O}{1}(t) \ket{\psi_g}
\end{align}
\end{subequations}
where $\ket{\psi_g}$ denotes the ground state of the
system Hamiltonian $\ophat{H}{}$, $\zeta = -1$ (resp. +1) for fermionic (resp. bosonic) operators. 
We will choose the ground state energy $E_g=0$ without loss of generality.
Typical choices of the operators are $\ophat{c}{\sigma},\ophat{c}{\sigma}^{\dagger}$, the creation
and annihilation operators, or $\ophat{S}{}^{-}, \ophat{S}{}^{+}$, the spin ladder operators for dynamical spin structure factors.

\subsection{Time evolution in the complex plane}\label{sec::complex_time_evolution}

\begin{figure}
\begin{tikzpicture}[scale=3.0]
\draw[->, very thick] (0,0) -- ++(0.65*1.5,0)node[yshift=-10]{$\mathrm{Re}(z)$}-- ++(0.65*1.5,0);
\draw[->, very thick] (0,0) -- ++(0,0.7)node[xshift=-20]{$-\mathrm{Im}(z)$}-- ++(0,0.7);
\draw[ultra thick,magenta] (0,0) -- (0,1.3);
\draw[ultra thick,cyan] (0,0) -- (1.3,0);
\draw[very thick] (0,0) .. controls (0.15,0.05*4) and (0.35,0.055*4) ..  (1.28,0.06*4);
\draw[very thick] (0,0) .. controls (0.15,0.05*6) and (0.35,0.055*6) ..  (1.24,0.06*6);
\draw[very thick] (0,0) .. controls (0.15,0.05*8) and (0.35,0.055*8) ..  (1.2,0.06*8);
\draw[->,very thick] (1.5, 0.06*3) -- ++(0,0.06*3)node[xshift=+10pt]{$\alpha_0$} -- ++ (0, 0.06*4);
\draw (0.7,0.0)node[yshift=+5pt,cyan, scale=0.8]{Real Time};
\draw (0.0,0.7)node[xshift=+5pt,magenta, scale=0.8, rotate=90]{Imaginary Time};
\end{tikzpicture}
\caption{Illustration of the time contours in the complex plane, cf Text.\label{fig::contour_illustration}}
\end{figure}
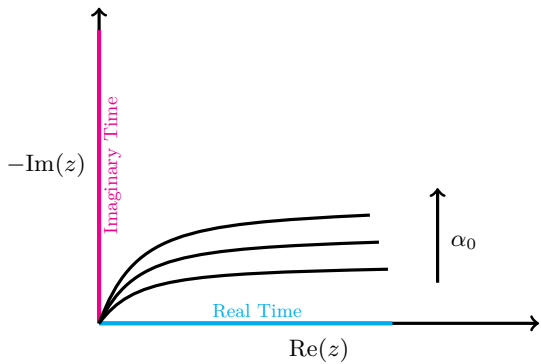

The first step of the method consists in evolving on a time contour in the complex plane.
Let us start with the parameterization of this contour, given by 
\begin{align}\label{eq::tau}
z(t,\alpha_0)\equiv\int_0^te^{-i\alpha_0 f(t')}dt',
\end{align}
where $t$ is real, $\alpha_0 \in [0, \frac{\pi}{2}]$, 
and $f$ is a real-valued smooth function such that $0\leq f(t) \leq 1$.
For the results discussed in this paper, we choose $f(t)=e^{-\frac{t}{2\pi}}$, 
even though our approach is general and not limited to this specific form.
Real and imaginary-time evolutions correspond to \mbox{$\alpha_0 = 0$} and \mbox{$\alpha_0 f(t) = \pi/2$},
respectively. 
This time contour is sketched in Fig.~\ref{fig::contour_illustration}. 
Crucially, we have the property $\text{Im} z(t, \alpha_0) < 0$.

The greater time-dependent correlation function of operators $ \ophat{O}{1},  \ophat{O}{2}$ can be generalized to the complex plane
\begin{align}\label{eq::gt_complex}
&G^{>}_{\ophat{O}{1} \ophat{O}{2}}(t, \alpha_0) \equiv -i \bra{\psi_g} \ophat{O}{1}
e^{-iz(t,\alpha_0)\ophat{H}{}}
\ophat{O}{2} \ket{\psi_g}, 
\end{align}
where $\ket{\psi_g}$ is the many-body ground state and $\ophat{H}{}$ is the Hamiltonian with the ground state energy $E_{g}$ subtracted
\footnote{In general, we want to use $\ophat{H}{}- E_0$, where $E_0$ is the ground state energy for a gapless system considered in this work, 
but we may want to use a different energy for a gapped system to ensure a better convergence.}.
The $\alpha_0$ dependence of $G^{>}(t,\alpha_0)$ is such that for $\alpha_0= 0$, $z(t, 0) = t$ and $G^{>}(t,0)= G^{>}(t)$.

The evaluation of $G^{>}$ involves the time-evolved state 
\begin{align}
\ket{\psi(t,\alpha_0)}\equiv e^{-iz(t,\alpha_0) \ophat{H}{} }\ophat{O}{2}\ket{\psi_g}
\end{align} 
for a time interval $t\in[0,t_{\text{max}}]$.
Due to the finite imaginary part $\mathrm{Im}z(t, \alpha_0) < 0$, 
we expect that the complex-time evolution of $\ket{\psi(t,\alpha_0)}$
gradually projects the state to the
low-energy manifold, and therefore that it will have lower entanglement than its real-time evolved counterpart
$\ket{\psi(t,0)}$. In practice, its MPS representation will require a
smaller bond dimension $\chi$ to reach the same accuracy. 
Purely imaginary time evolution is a limiting case, which exhibits nearly constant entanglement~\cite{Wolf_imaginary_time_2015} and
approaches the ground state at large time $t$.
The specific form of the contour chosen in this paper is designed to approximately
project into the low-energy manifold after a short time period, then remain in near-unitary evolution afterwards.
More detailed discussions on
complex time contours can be found in Appendix~\ref{app::contour_and_kernel}.

For the lesser Green's function we define an analogous complex time generalization 
\begin{align}
G^{<}_{\ophat{O}{1} \ophat{O}{2}}(t,\alpha_0) = -\zeta i \bra{\psi_g}\ophat{O}{2}e^{iz(t,\alpha_0)\ophat{H}{}}\ophat{O}{1} \ket{\psi_g}
\end{align}
with one important change: to suppress high-energy excitations one should employ a complex time contour with an imaginary part having the opposite sign, obtained by reversing the sign of the angle $\alpha_0\rightarrow -\alpha_0$, i.e., $z(t,\alpha_0)=\int_0^t e^{+ i\alpha_0 f(t')}dt'$.
Below we will primarily discuss $G^{>}(t,\alpha_0)$ but an otherwise similar analysis follows for $G^{<}(t,\alpha_0)$.
For brevity, we omit the operator indices in $G^{>}_{\ophat{O}{1} \ophat{O}{2}}(t,\alpha_0)$ when no confusion arises.

\subsection{Reconstruction of the correlation function on the real axis}\label{sec:reconstruct_real_axis}

The second step of the method consists in reconstructing the function on the real axis
from the complex time evolved function $G^{>}(t,\alpha_0)$.
The crucial balance of our approach is to go far enough in the complex plane to reduce 
the rank (or bond dimension) of the MPS, but to stay close enough to the real axis to reconstruct the final result efficiently and to high precision.

A natural possibility would be to use an analytical continuation technique, 
such as MaxEnt~\cite{Kai_complex_time_fciqmc_2018}, as $G^{>}$ is related to the spectral function 
as discussed in Appendix~\ref{app::contour_and_kernel}.
However, because analytic continuation is in general an ill-conditioned inversion problem, 
we prefer to use a more controlled reconstruction technique in this paper.

We consider the perturbative expansion of $G^{>}$ in powers of 
$ (-\alpha_0)$ {\it starting from the complex contour} $\alpha_0 > 0$ 
and evaluate it on the real axis $\alpha_0=0$.
Formally, we have
\begin{align}\label{eq::expansion}
G^{>}(t, 0) &= G^{>}(t, \alpha_0) + \sum_{n\geq 1} \frac{(-\alpha_0)^n}{n!}
\partial_{\alpha_0}^n  G^{>}(t, \alpha_0)
\end{align}
%
The terms of this series can be straightforwardly computed from the auxiliary quantities,
\begin{align}
\phi^{(n)}(t, \alpha_0) \equiv \bra{\psi_g}\ophat{O}{1} \ophat{H}{}^n \ket{\psi(t,\alpha_0)},
\end{align}
The relation between the perturbative series and $\phi^{(n)}$ can be explicitly derived from a simple chain rule, 
with more details given in Appendix~\ref{app::expansion}.
The left hand side  $\ophat{H}{}^n\ophat{O}{1}^\dagger \ket{\psi_g}$
can be obtained by representing $\ophat{H}{}$ as an MPO and using MPO-MPS multiplication.

In this expansion, we expect the zero-th order term $G^{>}(t,\alpha_0)$ to mainly capture the 
low-energy features of the spectra. The
higher-order terms $\partial_{\alpha_0}^n G^{>}(t,\alpha_0)$, which are linear combinations of
$\phi^{(l \leq n)}(t, \alpha_0)$, contain transitions between the ground state
$\ket{\psi_g}$ and higher-energy states generated by powers of the Hamiltonian
$\hat H$. We therefore expect those terms to reconstruct the high energy features of the spectral function.
In practice, one should choose $\alpha_0$ as small as possible, given a maximum acceptable bond dimension, 
to minimize the number of terms required to converge the Taylor series \eqref{eq::expansion}
and thereby the number of iterated powers of the Hamiltonian.

\section{Benchmark on the Single-Impurity Anderson Model}\label{sec::results_siam}

\subsection{Model}

In this section, we benchmark our approach on
the one-band single impurity Anderson model (SIAM).
Its Hamiltonian reads
\begin{equation}
\begin{split}
\ophat{H}{} & = \ophattext{H}{loc} + \ophattext{H}{bath} \\
\ophattext{H}{loc} & = \epsilon_d \sum_{\sigma = \uparrow, \downarrow} \ophat{n}{d\sigma} + U \ophat{n}{d\uparrow} \ophat{n}{d\downarrow}, \\
\ophattext{H}{bath} & = \sum_{b=0\atop\sigma  = \uparrow, \downarrow}^{N_b-1}\epsilon_b \ophat{n}{b\sigma} + \sum_{b=0\atop\sigma = \uparrow, \downarrow}^{N_b-1} \left(v_b \ophat{c}{b\sigma}^{\dagger}\ophat{d}{\sigma} + \mathrm{h.c.} \right)
\end{split}
\end{equation}
where $\ophat{d}{\sigma}$ and $\ophat{c}{b\sigma}$ are respectively the electron annihilation operators at the impurity site and bath site $b$ with spin $\sigma=\uparrow, \downarrow$. The corresponding density operators are $\ophat{n}{d\sigma}$ and $\ophat{n}{b\sigma}$. The bath parameters $\{\epsilon_b, v_{b}\}$ are determined by uniformly discretizing a hybridization function $\Delta(\omega)$ with a semielliptic spectrum $-\frac{1}{\pi}\mathrm{Im}\Delta(\omega)=\frac{2}{\pi D}\sqrt{1-\left( \frac{\omega}{D}\right)^2 }$ of half bandwidth $D$ into $N_b$ intervals.
We use the ``natural orbital'' basis, which was shown to greatly reduce the required computational cost~\cite{Lu2014,Lu2019,cao_tree_imp_2021}.

We compute the retarded Green's function at zero temperature
\begin{align}\label{eq::gretard}
    &G_{\sigma\sigma'}^R(t)  = -i \uptheta(t) \bra{\psi_g}\{\ophat{d}{\sigma}(t),\ophat{d}{\sigma'}^\dagger\} \ket{\psi_g},
\end{align}
Results presented in this section are obtained using the two-site TDVP for time
evolution with time step $dt=0.1/D$ and truncation weight $t_w=10^{-11}$. To
generate $H^n \ket{\psi(t, \alpha_0)}$, we use the zip-up algorithm to apply
$\hat{H}^n$~\cite{Stoudenmire_metts_2010}, setting a maximum bond dimension of
\mbox{$\chi_{max}=1000$}. All computations are conducted while preserving the global
$U(1)_{\text{charge}}$ and $U(1)_{S_z}$ symmetries. Unless otherwise noted, the
complex time evolution is performed with $\alpha_0=0.1$, and the spectra are
reconstructed up to order $n=4$. The NRG data presented in this work was obtained in a state-of-the-art implementation~\cite{nrg_fabian_1,nrg_fabian_2,nrg_fabian_3,nrg_fabian_4,nrg_fabian_5,nrg_fabian_6} based on the QSpace tensor library~\cite{nrg_fabian_7}, using a symmetric improved estimator for the self-energy~\cite{fabian2022}.

\begin{figure}[t]
	\includegraphics[width=\columnwidth]{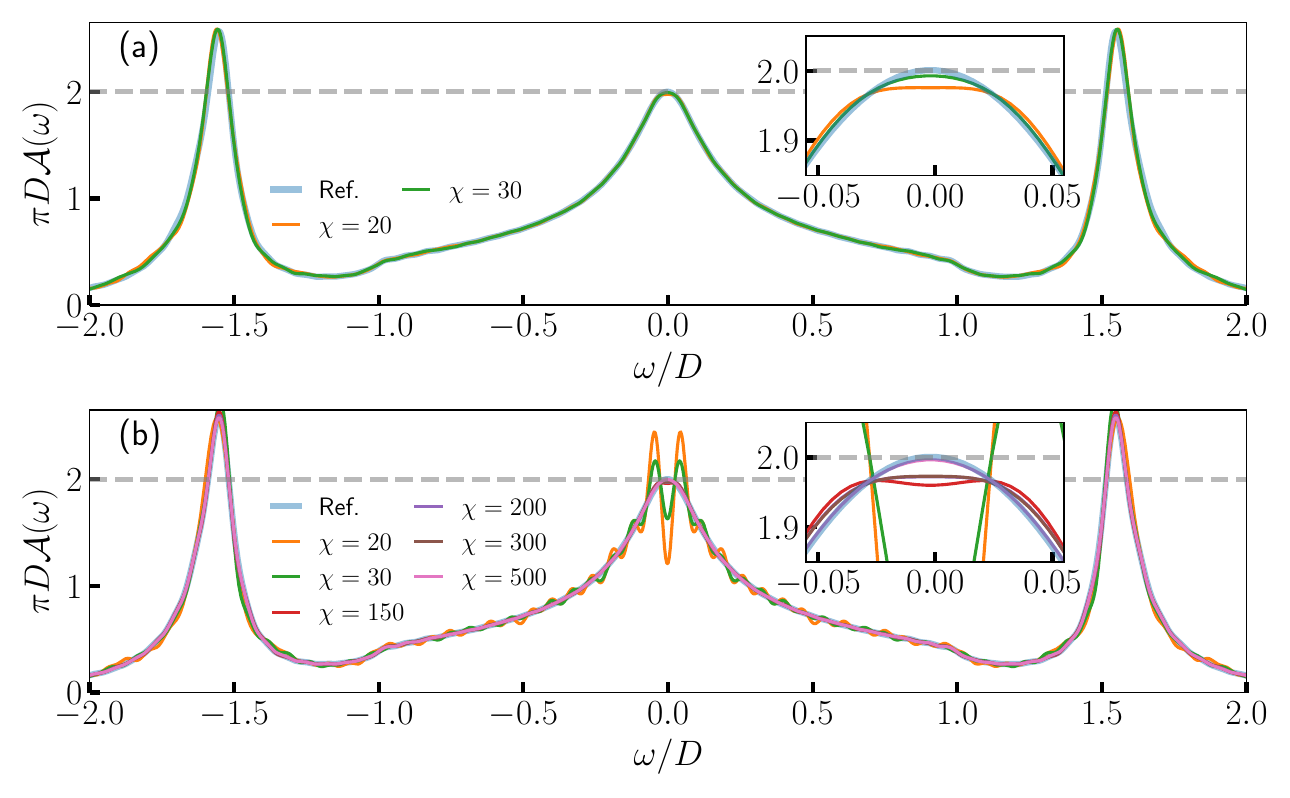}
	\caption{ Spectral functions obtained from (a) complex and (b) real
	   time evolution with different bond dimensions $\chi$.  The spectral function from real-time
	   evolution with $\chi=700$ is shown as a reference in both panels.
	   The complex time evolution is performed with $\alpha_0=0.1$ and the
	   spectra expanded up to order $n=4$. For all spectra, we set
	   $N_b=59$, $U=2D$, and $Dt_{\text{max}}=90$. Insets show zoom around
	   $\omega = 0$.
\label{fig::alpha_improment} }
\end{figure}

\subsection{Results}

From the retarded Green's function on the real-time axis, one can compute the spectral function
\begin{align}\label{eq::spectral_function}
\mathcal{A}_{\sigma\sigma'}(\omega) = -\frac{1}{\pi} \mathrm{Im} \int_{0}^{+\infty}dt\, e^{it\omega}\, G_{\sigma\sigma'}^R(t) \ .
\end{align}
We show the spectral function obtained by the complex time evolution and series extrapolation, 
along with ordinary real time evolution, on Fig.~\ref{fig::alpha_improment}.
As a reference, we show the result of real-time evolution with a large
bond dimension $\chi=700$.
The complex-time result reproduces the entire spectrum with bond dimensions $\chi \sim 20$, including
both the low-energy Kondo resonance and the high-energy Hubbard satellite structures.
In particular, the Friedel sum rule~\cite{Luttinger_1960,Luttinger_1961}, which dictates that
$\mathcal{A}(0)=2/(\pi D)$, is satisfied within an error of less than $0.3\%$
for $\chi=30$. 
In sharp contrast, the pure real time evolution require a much larger $\chi\sim 500$ 
to converge, especially for low-energy properties.

\begin{figure}[t]
    \includegraphics[width=0.5\textwidth]{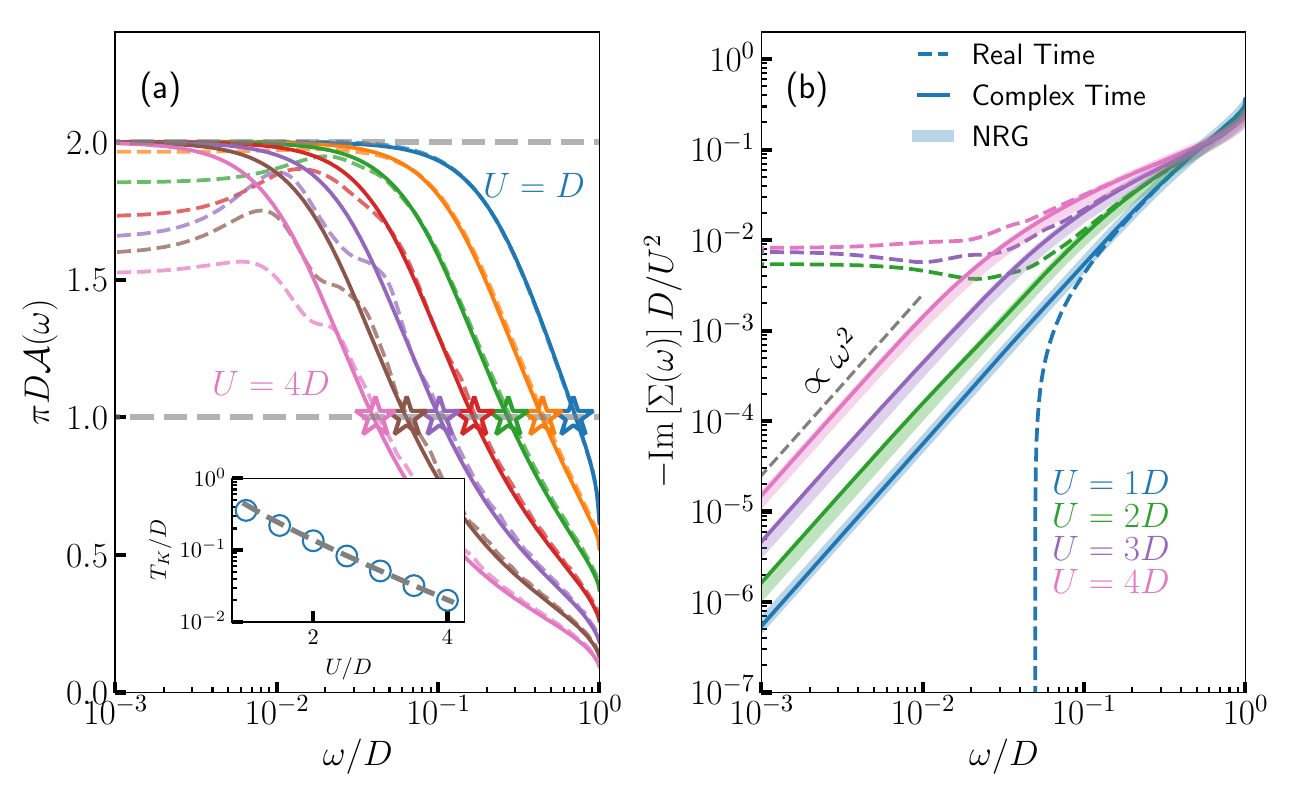}
    \caption{Low-energy (a) spectra and (b) self-energy obtained from complex
       (solid lines) and real (dashed lines) time evolutions for $U=D$ to $4D$.
       The inset in (a) shows the Kondo energy scale (blue open circles)
       extracted from complex-time spectra (cf text), and its fitting to the
       Schrieffer-Wolff limit (grey dashed line). Stars in panel {\it a} indicate the Kondo energy determined from DDMRG~\cite{Raas_siam_kondo_2005}. As a reference, the NRG
       self-energies are plotted (transparent lines) in
       panel {\it b}~\cite{fabian2022}. All results are calculated with $N_b=399$,
       $\chi=80$ and $Dt_{\text{max}}=240 (480)$ for $U\leq 2D$ ($U>2D$). A
       Lorentzian broadening with $\eta=0.01D$ is applied during the Fourier
       transformation of real-time data, while no broadening is used for the
       complex-time data. 
    \label{fig::kondo} }
\end{figure}

Let us now focus on the low energy part of the spectrum, for various $U$.
Figure~\ref{fig::kondo} shows the evolution of the spectral function and
self-energy at low frequency with increasing $U$ values.
For $U>D$, the real-time results show strong deviations around
$\omega=0$, which increase with $U$ (in particular a violation of Friedel sum rule and a lost of spectral weight).
On the other hand, the complex-time results satisfy the Friedel sum rule 
for all $U$ values.
Furthermore, we can extract the Kondo energy scale $k_BT_K$, 
defined here as the half width at half maximum 
of the
Kondo peak. It decreases exponentially with $U$ as expected in the  Schrieffer-Wolff
limit~\cite{hewson1997kondo}, with $T_K/D \approx
1.88\sqrt{D/U}\exp(-\pi\frac{UD}{16})$,
as shown in the inset of Fig.~\ref{fig::kondo}(a). 
It is also in excellent agreement with previous results obtained using
DDMRG~\cite{Raas_siam_kondo_2005}, as shown by the frequency values labeled by stars in  Fig.~\ref{fig::kondo}(a).
Finally, the Fermi liquid low energy behaviour $-\mathrm{Im}\Sigma(\omega)\sim (\omega/D)^2$
is recovered, with a high precision agreement with the NRG results~\cite{fabian2022}. 
In contrast, purely real time evolution is unable to obtain the Fermi liquid self-energy to such a high precision.

\begin{figure}[t]
	\includegraphics[width=0.49\textwidth]{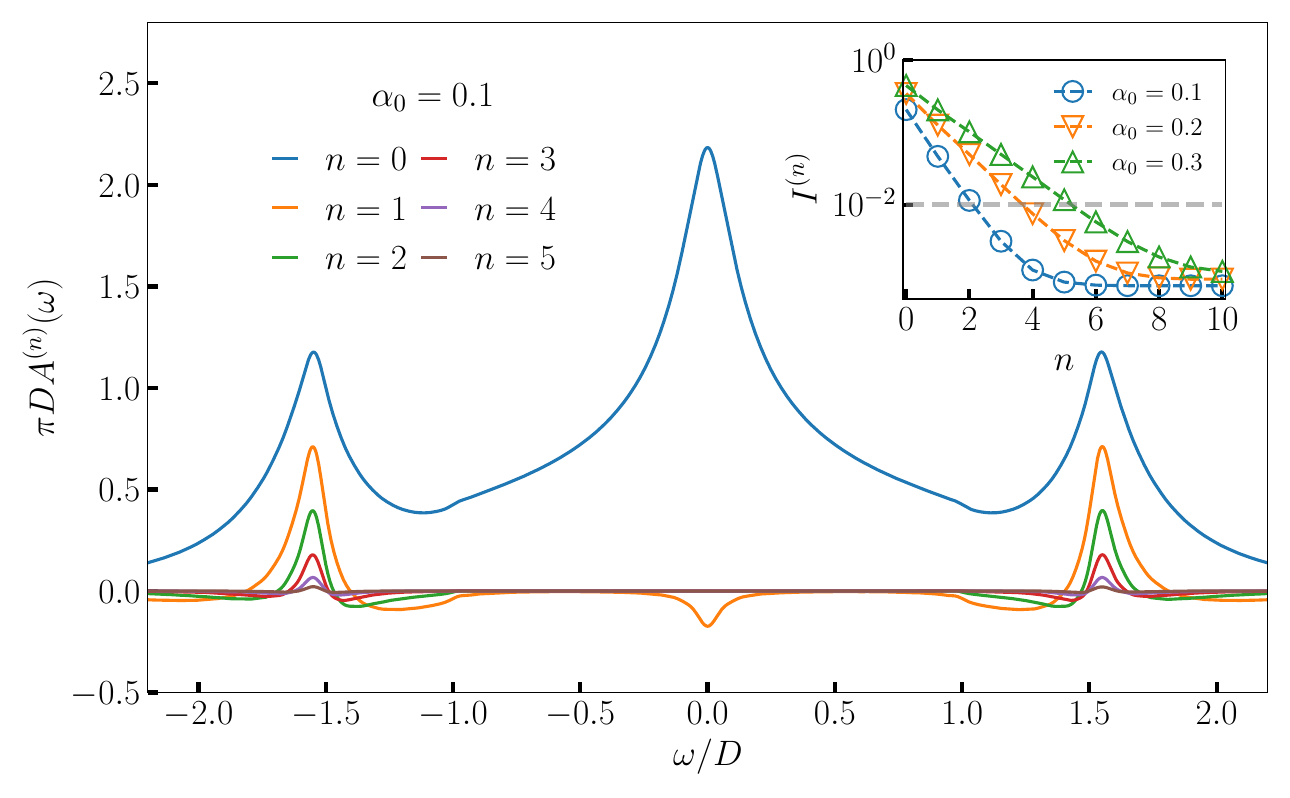}
	\caption{\label{fig::order_convergence} Contribution from each expansion order to the spectral function, $A^{(n)}(\omega)$. The insert shows the cumulative spectral weight error $I^{(n)}$ up to order $n$ (see main text) for different $\alpha_0$ values. $N_b=59$, $U=2D$, $\chi=500$, and $Dt_{\text{max}}=90$.
 }
\end{figure}

\subsection{Discussion}

Let us now study more in details the contribution of the various terms in the
Taylor expansion \eqref{eq::expansion} to the spectral function:
\begin{equation} 
\mathcal{A}(\omega)=\sum_{n=0}A^{(n)}(\omega) 
\end{equation}
In Figure~\ref{fig::order_convergence}, we show $A^{(n)}(\omega)$ up to $n=5$
for $\alpha_0=0.1$.  We observe that $A^{(0)}(\omega)$ gives the largest
contribution, especially for small $\omega$, followed by the first-order
expansion $A^{(1)}(\omega)$.  For $n\geq2$, $A^{(n)}(\omega)$ contributes
mainly to the high energy features.
A closer look at the contributions around
the Hubbard bands $\omega/D\sim \pm 1.55$ reveals a monotonic decrease in
$A^{(n)}(\omega)$ as the expansion order $n$ increases. 

The convergence of the Taylor series naturally depends on the value of $\alpha_0$.
The larger is $\alpha_0$, the more orders are needed.
Note that the radius of convergence of the series also depends on $\alpha_0$.
This is illustrated in the inset in Fig.~\ref{fig::order_convergence}, which 
shows the difference between the cumulative spectral function up to the expansion order
$n$ and the reference data used in Fig. \ref{fig::alpha_improment}:
\begin{equation} 
I^{(n)}=\int d\omega\
\left| \sum_{m=0}^n A^{(m)}(\omega) - \mathcal{A}^{\text{ref}}(\omega) \right|
\end{equation}
In order to achieve a precision of $10^{-2}$, we need 
$n=2$ for $\alpha_0=0.1$, $n=4$ for $\alpha_0=0.2$ and $n=5$ for $\alpha_0=0.3$.

\begin{figure}[ht]
	\includegraphics[width=0.49\textwidth]{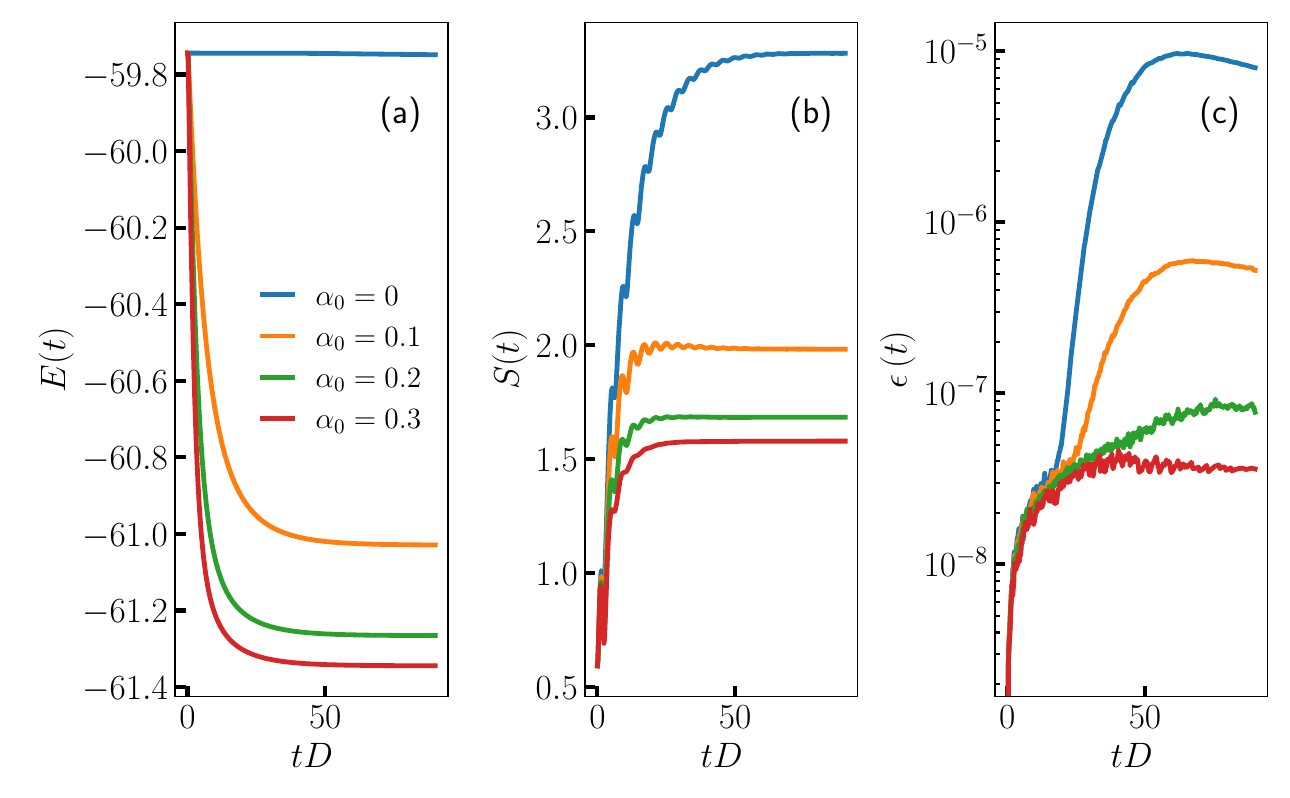}
	\caption{\label{fig::E_S_Error}(a) Energy, (b) entanglement entropy between the impurity and first bath site, and (c) two-site TDVP truncation error (at fixed $\chi=500$) of the time evolved state $\ket{\psi(t,\alpha_0)}$ as a function of time $t$ for various $\alpha_0$ values. The initial state is $\ket{\psi(0, \alpha_0)}=d^\dagger_\uparrow\ket{\psi_0}$. $N_b=59$ and $U=2D$.}
\end{figure}

The effect of the complex time evolution is further documented on 
Figure~\ref{fig::E_S_Error}.
On panel {\it a)}, we present the  
energy $E(t)=\bra{\psi(t,\alpha_0)}
\ophat{H}{}\ket{\psi(t,\alpha_0)}/\braket{\psi(t,\alpha_0)}{\psi(t,\alpha_0)}$.
As expected, for $\alpha_0=0$ it remains constant, but decreases with $t$ for $\alpha_0>0$, faster for larger $\alpha_0$, 
as the complex time evolution projects to the low energy manifold.
On Fig.~\ref{fig::E_S_Error}(b), we present the entanglement entropy between the impurity and the first bath site $S(t)$.
Its growth at long time is strongly reduced by $\alpha_0$.
 This typically results in a smaller MPS
bond dimension $\chi$ when working on the same level of accuracy.
Consistently, the two-site TDVP truncation error $\epsilon$ shown in
Fig.~\ref{fig::E_S_Error}(c) is significantly reduced at large
$\alpha_0$. These results illustrate the fact that 
complex time evolution of $\ket{\psi(t,\alpha_0)}$
is easier to compute with MPS (i.e. require a lower bond dimension $\chi$).

Regarding the computational cost of our method, as with real-time evolution, the most computationally demanding part is to compute the time-evolved state $\ket{\psi(t,\alpha_0)}$ over the time range $[0, t_{\text{max}}]$. The cost of MPS-based time-evolution methods, such as TDVP used in this work, scales with MPS rank or bond dimension as $\chi^3$.  We find that to reach the same accuracy per time step, a much smaller $\chi$ is needed along the complex contour versus real time. Additionally, the number of Krylov steps needed inside the TDVP algorithm is fewer. Finally, for a small enough $\alpha_0\sim0.1$, the Taylor series reconstruction of the real-time results 
converges rapidly with just a few orders (i.e., $n\leq4$).
See Appendix~\ref{app::computational_cost}) for further discussion of these points.

\section{Conclusion}\label{sec::conclusion} 

In summary, the combination of a complex time evolution of tensor network states
and a perturbative reconstruction of the real time result
is an efficient approach to compute the spectral function on real frequencies.
We have shown this approach can deliver very high-precision results at low energy, comparable to NRG,
using detailed benchmarks for the one band SIAM model.
The complex time evolution succeeds in capturing the whole spectrum
of the model, including the exponentially small Kondo and Fermi liquid energy
scale, while real-time evolution struggles in the long-time dynamics, hence at low frequency, 
and requires a relatively large broadening.

Our approach can be extended to various tensor network
states~\cite{Daniel_fork_2017,Kloss_tree_2020,cao_tree_imp_2021} where
efficient time evolution is feasible.
When applied to complicated systems characterized by an MPO representation of the Hamiltonian with 
large bond dimension, efficiently computing the higher-order expansion
coefficients could however become challenging. 
We emphasize however that the Taylor series expansion presented here is only one possibility
of extrapolating the function from finite $\alpha_0$ to $\alpha_0=0$. Another possibility could 
be extrapolating from a few finite values of $\alpha_0$ down to $\alpha_0 = 0$.

Recently there have been a number of new methods proposed to efficiently
access long-time dynamics using tensor networks. The strategies used range from novel
truncations of the quantum state \cite{White18,Ye2021,FriasPerez}, compression of the process tensor or influence functional 
capturing bath dynamics \cite{PhysRevLett.102.240603,Muller-Hermes_2012,PhysRevB.89.201102, Strathearn2018,Fux21,Lerose21,Thoenniss_Nonequlibrium,Thoenniss_Efficient,Ng23,Kloss23}, compression of high-order perturbative series \cite{PhysRevX.12.041018,Erpenbeck_2023}, or controllably introducing dissipation 
\cite{Rakovszky22, Keyserlingk21, Azad}.
While our approach has some conceptual similarity with the dissipation approaches, it has important practical advantages. 
It is straightforward to implement for any type of system, does not rely on special bath properties,
offers controlled schemes to reconstruct the real-time dynamics, and delivers high-precision results.
Moreover, it could be blended with the other methods above.

We leave for future work the question of multi-orbital quantum impurity models.
A key question is whether complex time evolution will reach a similar accuracy 
in the low frequency self-energy for larger systems. This would open a route
to a high precision real-frequency solutions for five and more orbitals, which are 
currently out of reach of the NRG algorithm.

\begin{acknowledgments}
We thank Fabian B. Kugler for providing the NRG data. We also thank Antoine Georges, Benedikt Kloss, Kun Chen, Martin Grundner, and Ulrich Schollw\"ock for insightful discussions. 
The Flatiron Institute is a division of the Simons Foundation. 
Y. L. acknowledges support from the National Key R\&D Program of China (No. 2022YFA1403000) and the National Natural Science Foundation of China (No. 12274207).
\end{acknowledgments}

\appendix
\counterwithin{figure}{section}

\section{Complex time contour and associated kernel}\label{app::contour_and_kernel}

\begin{figure}[h!]
\includegraphics[width=0.49\textwidth]{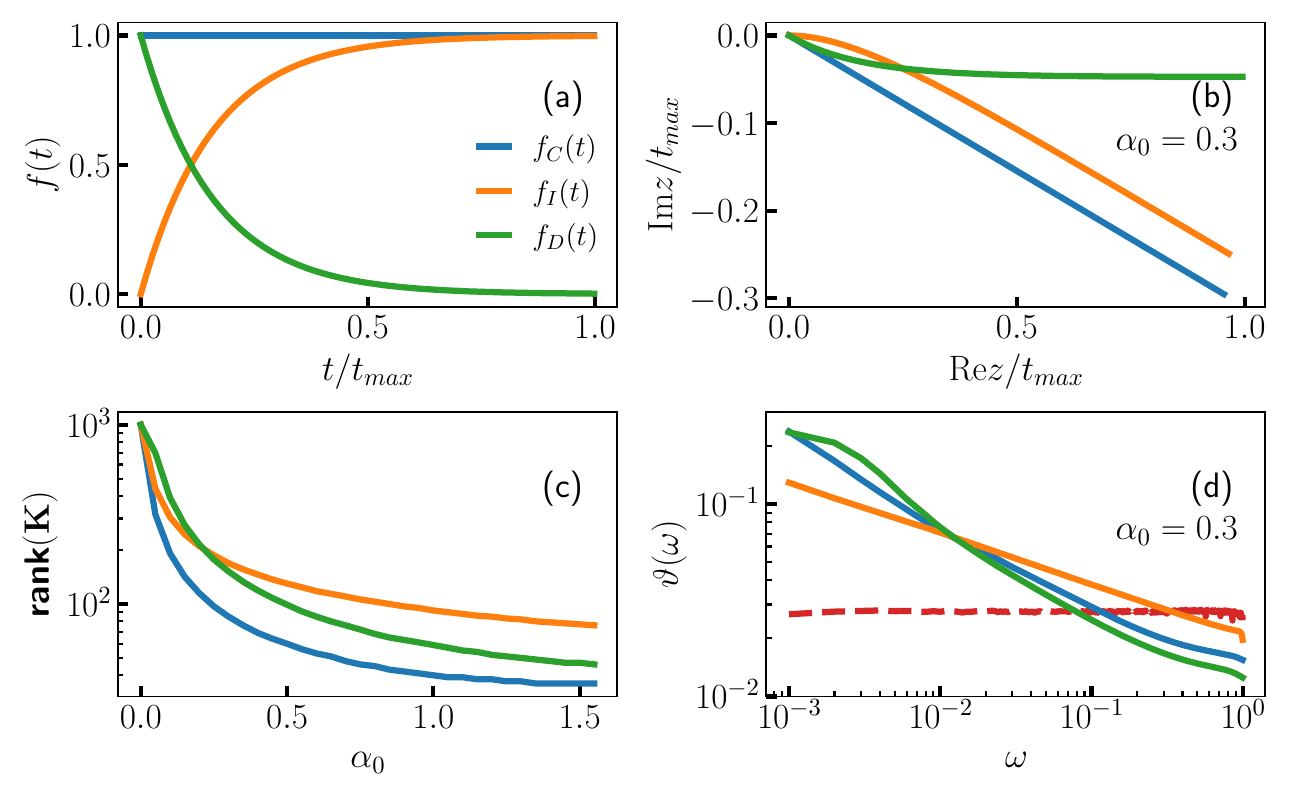}
\caption{\label{fig::kernel} Complex time contour and associated kernel properties. (a) Illustration of three exemplary complex time contours: constant angle contour $f_C(t)$ (blue), decreasing angle contour $f_D(t)$ (green), and increasing angle contour $f_I(t)$ (orange). (b) The exemplary three contours in the complex time plane, with $\alpha_0=0.3$. (c) Kernel rank as a function of $\alpha_0$ for the illustrated contours. For our analysis, we opt a linear grid with $\omega\in\left[0,1\right]$ has $N_w=1000$ points, and $t\in\left[0,t_{\text{max}}\right]$ has $N_t=1000$ points. The time interval is set to $t_{\text{max}}=2\pi/d\omega$ to accommodate the smallest energy scale $d\omega=1/N_w$. (d) $\vartheta(\omega)$ indicating the resolution of $\vecbf{K}$ on the real frequency $\omega$ for the three exemplary contours with $\alpha_0=0.3$. The uniform resolution from real-time evolution is also plotted for reference (dash red line). All panels use the same color scheme for the contours as indicated in panel (a).}
\end{figure}

The Green function in complex time $G^{>}(t, \alpha_0)$ and its spectral function $\mathcal{A}^{>}(\omega)$ are related by 
\begin{align}\label{eq::Aw_complex_fourier}
&G^{>}(t, \alpha_0) = \int d\omega \mathcal{A}^{>}(\omega) e^{-i z(t,\alpha_0) \omega}.
\end{align}
However, computing $\mathcal{A}$ from $G^{>}$ is an ill-conditioned problem for finite $\alpha(t)$. To be concrete, consider discretizing Eq.~\eqref{eq::Aw_complex_fourier} uniformly with $N_t$ time points and $N_w$ frequency points. The kernel connecting $G(t, \alpha_0)$ and $\mathcal{A}(\omega)$ can now be interpreted in a matrix-vector format as
\begin{gather}
	\vecbf{G} = \vecbf{K}\cdot \vecbf{A}, 
\end{gather}
where each kernel matrix element is simply given by $\vecbf{K}_{ij} =e^{-i z_i \omega_j} d\omega$. Let us consider three exemplary complex time contours $f(t)$ as depicted in Fig.~\ref{fig::kernel}(a): a contour with a constant angle $f_C(t)=1$(blue line), a contour with increasing angle $f_I(t)=1 - e^{-t\omega_0}$ (orange line), and a contour with decreasing angle $f_D(t)=e^{-t\omega_0}$ (green line). Here, $\omega_0=2\pi/t_{\text{max}}$ is the lowest energy one can resolve for a time duration up to $t_{\text{max}}$. Fig.~\ref{fig::kernel}(b) shows their corresponding contour in the complex time plane with $\alpha_0=0.3$. As depicted in Fig.~\ref{fig::kernel}(c), for all contours, the kernel rank $\textbf{rank}(\vecbf{K})$ diminishes rapidly with the growth of $\alpha_0$. The difficulty of analytical continuation from complex time to real frequency (including the more commonly encountered case of imaginary time to real frequency) arises from the reduced rank of $\vecbf{K}$. This ambiguity can be alleviated by choosing a small $\alpha_0$, e.g., positioning the complex contour near the real axis.

The kernel rank does not directly reveal the resolution on the real frequency axis. A more quantitative understanding can be drawn from the Singular Value Decomposition (SVD) of the kernel as $\vecbf{K} = U\cdot S\cdot V^{\dagger}$. 
The columns of $V^\dagger$ constitute an orthogonal basis set along the frequency axis, each being weighed by its respective singular values in $S$. We define 
\begin{gather}
 \vartheta(\omega_i)=\frac{\sum_{r=1}^{\textbf{rank}{(\vecbf{K})}} S_r^2 |(V^\dagger)_{r\omega_i}|}{\sum_{r=1}^{\textbf{rank}{(\vecbf{K})}} S_r^2},
\end{gather}
as a measure of the kernel's resolution on the real frequency axis. As shown in Fig.~\ref{fig::kernel}(d), in contrast to the uniform resolution of real-time evolution (dash red line), all three contours with finite $\alpha_0$ values lean towards emphasizing the low-energy part. Notably, the contour $\alpha(t)=\alpha_0f_D(t)$ provides the most substantial emphasis on this region.

The core motivation behind leveraging complex time evolution in the MPS
framework is the gradual suppression of high-energy excitations throughout the
time evolution. This approach works better when long-time states are predominantly
characterized by a limited set of low-energy excitations, thereby improving its
MPS representation. Our primary goal is that
$G^{>}(t,\alpha_0)$ mainly encapsulates the low-energy physics. We therefore choose 
$f_D(t)$ as the contour for all the results presented in this paper.

\section{Expansion in $\alpha_0$}\label{app::expansion}
In this section, we elaborate on the Taylor expansion in Eq.~\eqref{eq::expansion}, and explicitly work out the first few expansion terms.
We define 
\begin{align}
    J^{(n)}(t,\alpha_0)&:=-i\partial^n_{\alpha_0} z(t,\alpha_0), \\ \nonumber
    &=-i\int_0^t\left[ -if(t') \right]^n e^{-i\alpha_0f(t')}dt',
\end{align}
and $g^{(n)}(t,\alpha_0):=\frac{(-\alpha_0)^n}{n!}\partial^n_{\alpha_0} G^{>}(t,\alpha_0)$. Then, the first four Taylor expansion terms are
\begin{widetext}
\begin{align}
    g^{(1)} &= -\alpha_0 J^{(1)}\phi^{(1)}, \\ \nonumber
    g^{(2)} &= 
    \frac{\alpha_0^2}{2!} \left[ J^{(2)}\phi^{(1)} + \left(J^{(1)}\right)^2\phi^{(2)}\right], \\ \nonumber
    g^{(3)} &= 
    \frac{-\alpha_0^3}{3!} \left[ J^{(3)}\phi^{(1)} + 3J^{(1)}J^{(2)}\phi^{(2)}+\left(J^{(1)}\right)^3\phi^{(3)}\right], \\ \nonumber
    g^{(4)} &= 
    \frac{\alpha_0^4}{4!} \left[ J^{(4)}\phi^{(1)} + \left( 4J^{(1)}J^{(3)} + 3\left(J^{(2)}\right)^2\right)\phi^{(2)} 
    + 6\left(J^{(1)}\right)^2 J^{(2)} \phi^{(3)}
    +\left(J^{(1)}\right)^4\phi^{(4)}\right].
\end{align}
\end{widetext}
Here, we omit the explicit dependence on $t$ and $\alpha_0$ in $J^{(n)}(t,\alpha_0), \phi^{(n)}(t,\alpha_0)$ and $g^{(n)}(t,\alpha_0)$ for brevity.

\section{Computational Cost}\label{app::computational_cost}

\begin{figure}[h!]
\includegraphics[width=0.49\textwidth]{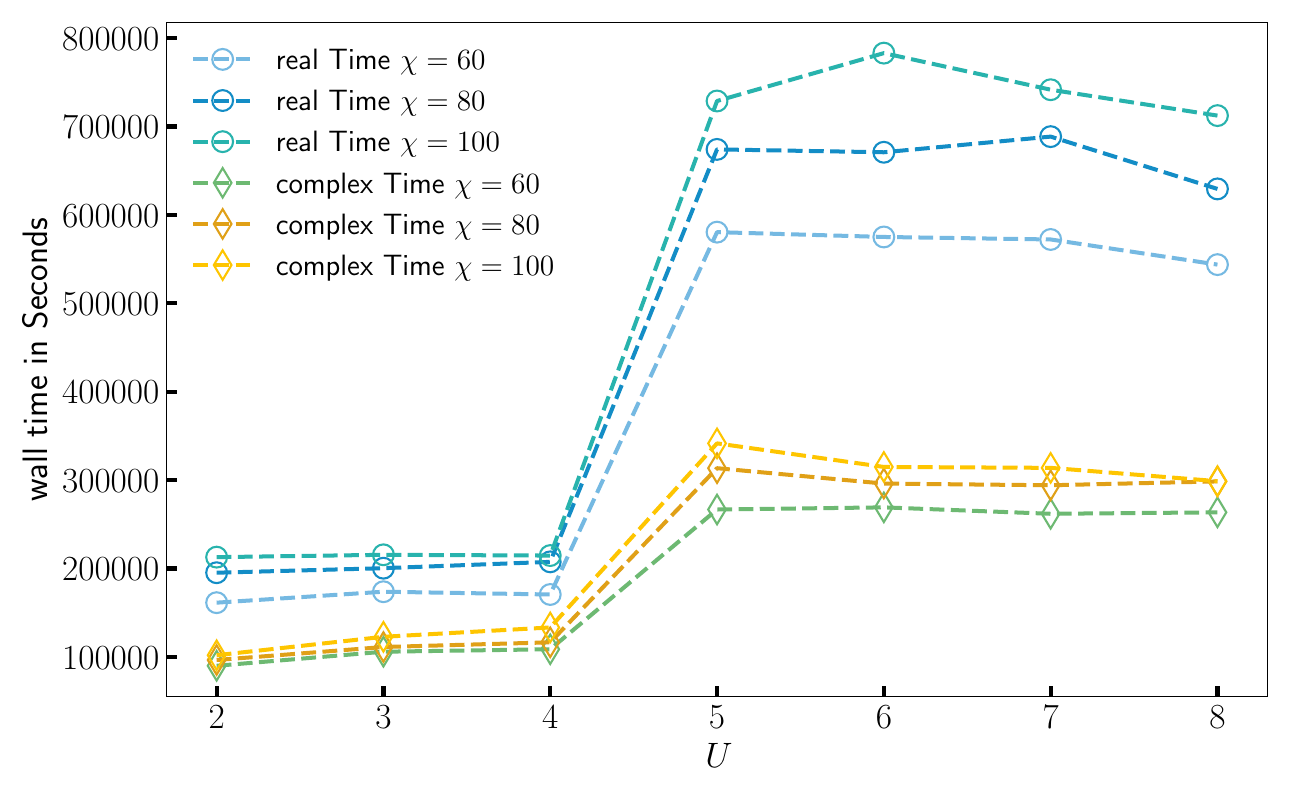}
\caption{\label{fig::computational_time} Comparison of actual or ``wall'' time for complex and real-time evolution with various interaction strengths and bond dimensions. Wall time for ``complex time'' includes a full implementation of our method. For $U\leq 2D$, we have $t_{\text{max}}=60D$ and $t_{\text{max}}=120D$ for $U>2D$. }
\end{figure}

For the proposed method, when employing two-site TDVP~\cite{tdvp_rk_2011, tdvp_local_2016} for time evolution and the zip-up algorithm to apply the Hamiltonian~\cite{Stoudenmire_metts_2010}, the primary computational cost scales as $N_tn_kd^2k\chi^3 + nkd\chi'^3  + N_tnd\chi'^2\chi$. The first term stems from the two-site TDVP with maximal bond dimension $\chi$. The second term arises from applying the Hamiltonian MPO with bond dimension $k$ on the state with a maximum bond dimension of $\chi'$($\chi'\geq\chi$) $n$ times to generate $\ket{\phi^n}$. 
The third term arises from evaluating the overlap  $\langle\phi^n|\psi(t,\alpha_0)\rangle$. Here, $d$ represents the local Hilbert space size ($d=4$ for an electron site), $N_t$ is the number of time evolution steps, and $n_k$ is the average local Krylov space size required to exponentiate the local evolution operator($n_k\sim\mathcal{O}(10)$). 
For small $\alpha_0$, the Taylor series converges rapidly in the expansion order $n$ ($n\sim 4$ for SIAM with $\alpha_0=0.1$), and $N_t$ is typically very large ($N_t\sim \mathcal{O}(10^3)$). 
Thus, akin to real-time evolution, the most computationally demanding part is calculating the time-evolved state $\ket{\psi(t,\alpha_0)}$ on the complex time contour.

As discussed in Sec.~\ref{sec::complex_time_evolution} in the main text, when employing complex time evolution, high-energy excitations are progressively filtered out, allowing the state to be efficiently and accurately captured by an MPS over an extended time range. Therefore, in comparison to real-time evolution, a significantly smaller bond dimension $\chi$ can achieve comparable accuracy. Moreover, since states at long times closely resemble the ground state, the average local Krylov space size $n_k$ is also reduced relative to real-time evolution, further reducing computational costs. 
For SIAM calculations, we plot the wall time (actual time) of calculations with various interaction strengths and bond dimensions in Fig.~\ref{fig::computational_time}. 
We note that the complex time evolution (including the expansion step) is approximately two to three times faster than the real-time evolution when using identical $\chi$ and truncation error $t_w$.

\bibliographystyle{apsrev4-2}
\bibliography{dynamics.bib}

\begin{thebibliography}{93}%
\makeatletter
\providecommand \@ifxundefined [1]{%
 \@ifx{#1\undefined}
}%
\providecommand \@ifnum [1]{%
 \ifnum #1\expandafter \@firstoftwo
 \else \expandafter \@secondoftwo
 \fi
}%
\providecommand \@ifx [1]{%
 \ifx #1\expandafter \@firstoftwo
 \else \expandafter \@secondoftwo
 \fi
}%
\providecommand \natexlab [1]{#1}%
\providecommand \enquote  [1]{``#1''}%
\providecommand \bibnamefont  [1]{#1}%
\providecommand \bibfnamefont [1]{#1}%
\providecommand \citenamefont [1]{#1}%
\providecommand \href@noop [0]{\@secondoftwo}%
\providecommand \href [0]{\begingroup \@sanitize@url \@href}%
\providecommand \@href[1]{\@@startlink{#1}\@@href}%
\providecommand \@@href[1]{\endgroup#1\@@endlink}%
\providecommand \@sanitize@url [0]{\catcode `\\12\catcode `\$12\catcode
  `\&12\catcode `\#12\catcode `\^12\catcode `\_12\catcode `\%12\relax}%
\providecommand \@@startlink[1]{}%
\providecommand \@@endlink[0]{}%
\providecommand \url  [0]{\begingroup\@sanitize@url \@url }%
\providecommand \@url [1]{\endgroup\@href {#1}{\urlprefix }}%
\providecommand \urlprefix  [0]{URL }%
\providecommand \Eprint [0]{\href }%
\providecommand \doibase [0]{https://doi.org/}%
\providecommand \selectlanguage [0]{\@gobble}%
\providecommand \bibinfo  [0]{\@secondoftwo}%
\providecommand \bibfield  [0]{\@secondoftwo}%
\providecommand \translation [1]{[#1]}%
\providecommand \BibitemOpen [0]{}%
\providecommand \bibitemStop [0]{}%
\providecommand \bibitemNoStop [0]{.\EOS\space}%
\providecommand \EOS [0]{\spacefactor3000\relax}%
\providecommand \BibitemShut  [1]{\csname bibitem#1\endcsname}%
\let\auto@bib@innerbib\@empty
\bibitem [{\citenamefont {Damascelli}\ \emph {et~al.}(2003)\citenamefont
  {Damascelli}, \citenamefont {Hussain},\ and\ \citenamefont
  {Shen}}]{shen2003}%
  \BibitemOpen
  \bibfield  {author} {\bibinfo {author} {\bibfnamefont {A.}~\bibnamefont
  {Damascelli}}, \bibinfo {author} {\bibfnamefont {Z.}~\bibnamefont
  {Hussain}},\ and\ \bibinfo {author} {\bibfnamefont {Z.-X.}\ \bibnamefont
  {Shen}},\ }\href {https://doi.org/10.1103/RevModPhys.75.473} {\bibfield
  {journal} {\bibinfo  {journal} {Rev. Mod. Phys.}\ }\textbf {\bibinfo {volume}
  {75}},\ \bibinfo {pages} {473} (\bibinfo {year} {2003})}\BibitemShut
  {NoStop}%
\bibitem [{\citenamefont {Ament}\ \emph {et~al.}(2011)\citenamefont {Ament},
  \citenamefont {van Veenendaal}, \citenamefont {Devereaux}, \citenamefont
  {Hill},\ and\ \citenamefont {van~den Brink}}]{vanden2011}%
  \BibitemOpen
  \bibfield  {author} {\bibinfo {author} {\bibfnamefont {L.~J.~P.}\
  \bibnamefont {Ament}}, \bibinfo {author} {\bibfnamefont {M.}~\bibnamefont
  {van Veenendaal}}, \bibinfo {author} {\bibfnamefont {T.~P.}\ \bibnamefont
  {Devereaux}}, \bibinfo {author} {\bibfnamefont {J.~P.}\ \bibnamefont
  {Hill}},\ and\ \bibinfo {author} {\bibfnamefont {J.}~\bibnamefont {van~den
  Brink}},\ }\href {https://doi.org/10.1103/RevModPhys.83.705} {\bibfield
  {journal} {\bibinfo  {journal} {Rev. Mod. Phys.}\ }\textbf {\bibinfo {volume}
  {83}},\ \bibinfo {pages} {705} (\bibinfo {year} {2011})}\BibitemShut
  {NoStop}%
\bibitem [{\citenamefont {Georges}\ \emph {et~al.}(1996)\citenamefont
  {Georges}, \citenamefont {Kotliar}, \citenamefont {Krauth},\ and\
  \citenamefont {Rozenberg}}]{Georges1996}%
  \BibitemOpen
  \bibfield  {author} {\bibinfo {author} {\bibfnamefont {A.}~\bibnamefont
  {Georges}}, \bibinfo {author} {\bibfnamefont {G.}~\bibnamefont {Kotliar}},
  \bibinfo {author} {\bibfnamefont {W.}~\bibnamefont {Krauth}},\ and\ \bibinfo
  {author} {\bibfnamefont {M.~J.}\ \bibnamefont {Rozenberg}},\ }\href
  {https://doi.org/10.1103/RevModPhys.68.13} {\bibfield  {journal} {\bibinfo
  {journal} {Rev. Mod. Phys.}\ }\textbf {\bibinfo {volume} {68}},\ \bibinfo
  {pages} {13} (\bibinfo {year} {1996})}\BibitemShut {NoStop}%
\bibitem [{\citenamefont {Kotliar}\ \emph {et~al.}(2006)\citenamefont
  {Kotliar}, \citenamefont {Savrasov}, \citenamefont {Haule}, \citenamefont
  {Oudovenko}, \citenamefont {Parcollet},\ and\ \citenamefont
  {Marianetti}}]{Kotliar2006}%
  \BibitemOpen
  \bibfield  {author} {\bibinfo {author} {\bibfnamefont {G.}~\bibnamefont
  {Kotliar}}, \bibinfo {author} {\bibfnamefont {S.~Y.}\ \bibnamefont
  {Savrasov}}, \bibinfo {author} {\bibfnamefont {K.}~\bibnamefont {Haule}},
  \bibinfo {author} {\bibfnamefont {V.~S.}\ \bibnamefont {Oudovenko}}, \bibinfo
  {author} {\bibfnamefont {O.}~\bibnamefont {Parcollet}},\ and\ \bibinfo
  {author} {\bibfnamefont {C.~A.}\ \bibnamefont {Marianetti}},\ }\href
  {https://doi.org/10.1103/RevModPhys.78.865} {\bibfield  {journal} {\bibinfo
  {journal} {Rev. Mod. Phys.}\ }\textbf {\bibinfo {volume} {78}},\ \bibinfo
  {pages} {865} (\bibinfo {year} {2006})}\BibitemShut {NoStop}%
\bibitem [{\citenamefont {Fannes}\ \emph {et~al.}(1992)\citenamefont {Fannes},
  \citenamefont {Nachtergaele},\ and\ \citenamefont
  {Werner}}]{Fannes_fcs_1992}%
  \BibitemOpen
  \bibfield  {author} {\bibinfo {author} {\bibfnamefont {M.}~\bibnamefont
  {Fannes}}, \bibinfo {author} {\bibfnamefont {B.}~\bibnamefont
  {Nachtergaele}},\ and\ \bibinfo {author} {\bibfnamefont {R.~F.}\ \bibnamefont
  {Werner}},\ }\href@noop {} {\bibfield  {journal} {\bibinfo  {journal}
  {Communications in mathematical physics}\ }\textbf {\bibinfo {volume}
  {144}},\ \bibinfo {pages} {443} (\bibinfo {year} {1992})}\BibitemShut
  {NoStop}%
\bibitem [{\citenamefont {\"Ostlund}\ and\ \citenamefont
  {Rommer}(1995)}]{Ostlund_fcs_1995}%
  \BibitemOpen
  \bibfield  {author} {\bibinfo {author} {\bibfnamefont {S.}~\bibnamefont
  {\"Ostlund}}\ and\ \bibinfo {author} {\bibfnamefont {S.}~\bibnamefont
  {Rommer}},\ }\href {https://doi.org/10.1103/PhysRevLett.75.3537} {\bibfield
  {journal} {\bibinfo  {journal} {Phys. Rev. Lett.}\ }\textbf {\bibinfo
  {volume} {75}},\ \bibinfo {pages} {3537} (\bibinfo {year}
  {1995})}\BibitemShut {NoStop}%
\bibitem [{\citenamefont {Verstraete}\ \emph {et~al.}(2008)\citenamefont
  {Verstraete}, \citenamefont {Murg},\ and\ \citenamefont
  {Cirac}}]{Frank_mps_2008}%
  \BibitemOpen
  \bibfield  {author} {\bibinfo {author} {\bibfnamefont {F.}~\bibnamefont
  {Verstraete}}, \bibinfo {author} {\bibfnamefont {V.}~\bibnamefont {Murg}},\
  and\ \bibinfo {author} {\bibfnamefont {J.}~\bibnamefont {Cirac}},\ }\href
  {https://doi.org/10.1080/14789940801912366} {\bibfield  {journal} {\bibinfo
  {journal} {Advances in Physics}\ }\textbf {\bibinfo {volume} {57}},\ \bibinfo
  {pages} {143} (\bibinfo {year} {2008})}\BibitemShut {NoStop}%
\bibitem [{\citenamefont {White}(1992)}]{dmrg_white_1992}%
  \BibitemOpen
  \bibfield  {author} {\bibinfo {author} {\bibfnamefont {S.~R.}\ \bibnamefont
  {White}},\ }\href {https://doi.org/10.1103/PhysRevLett.69.2863} {\bibfield
  {journal} {\bibinfo  {journal} {Phys. Rev. Lett.}\ }\textbf {\bibinfo
  {volume} {69}},\ \bibinfo {pages} {2863} (\bibinfo {year}
  {1992})}\BibitemShut {NoStop}%
\bibitem [{\citenamefont {White}(1993)}]{dmrg_white_1993}%
  \BibitemOpen
  \bibfield  {author} {\bibinfo {author} {\bibfnamefont {S.~R.}\ \bibnamefont
  {White}},\ }\href {https://doi.org/10.1103/PhysRevB.48.10345} {\bibfield
  {journal} {\bibinfo  {journal} {Phys. Rev. B}\ }\textbf {\bibinfo {volume}
  {48}},\ \bibinfo {pages} {10345} (\bibinfo {year} {1993})}\BibitemShut
  {NoStop}%
\bibitem [{\citenamefont {Schollwöck}(2011)}]{Ulrich_review_2011}%
  \BibitemOpen
  \bibfield  {author} {\bibinfo {author} {\bibfnamefont {U.}~\bibnamefont
  {Schollwöck}},\ }\href
  {https://doi.org/https://doi.org/10.1016/j.aop.2010.09.012} {\bibfield
  {journal} {\bibinfo  {journal} {Annals of Physics}\ }\textbf {\bibinfo
  {volume} {326}},\ \bibinfo {pages} {96} (\bibinfo {year} {2011})}\BibitemShut
  {NoStop}%
\bibitem [{\citenamefont {Hallberg}(1995)}]{Hallberg_lanczos_1995}%
  \BibitemOpen
  \bibfield  {author} {\bibinfo {author} {\bibfnamefont {K.~A.}\ \bibnamefont
  {Hallberg}},\ }\href {https://doi.org/10.1103/PhysRevB.52.R9827} {\bibfield
  {journal} {\bibinfo  {journal} {Phys. Rev. B}\ }\textbf {\bibinfo {volume}
  {52}},\ \bibinfo {pages} {R9827} (\bibinfo {year} {1995})}\BibitemShut
  {NoStop}%
\bibitem [{\citenamefont {Dargel}\ \emph {et~al.}(2011)\citenamefont {Dargel},
  \citenamefont {Honecker}, \citenamefont {Peters}, \citenamefont {Noack},\
  and\ \citenamefont {Pruschke}}]{Dargel_lanczos_refined_2011}%
  \BibitemOpen
  \bibfield  {author} {\bibinfo {author} {\bibfnamefont {P.~E.}\ \bibnamefont
  {Dargel}}, \bibinfo {author} {\bibfnamefont {A.}~\bibnamefont {Honecker}},
  \bibinfo {author} {\bibfnamefont {R.}~\bibnamefont {Peters}}, \bibinfo
  {author} {\bibfnamefont {R.~M.}\ \bibnamefont {Noack}},\ and\ \bibinfo
  {author} {\bibfnamefont {T.}~\bibnamefont {Pruschke}},\ }\href
  {https://doi.org/10.1103/PhysRevB.83.161104} {\bibfield  {journal} {\bibinfo
  {journal} {Phys. Rev. B}\ }\textbf {\bibinfo {volume} {83}},\ \bibinfo
  {pages} {161104} (\bibinfo {year} {2011})}\BibitemShut {NoStop}%
\bibitem [{\citenamefont {Dargel}\ \emph
  {et~al.}(2012{\natexlab{a}})\citenamefont {Dargel}, \citenamefont
  {W\"ollert}, \citenamefont {Honecker}, \citenamefont {McCulloch},
  \citenamefont {Schollw\"ock},\ and\ \citenamefont
  {Pruschke}}]{Dargel_lanczos_refined_2012}%
  \BibitemOpen
  \bibfield  {author} {\bibinfo {author} {\bibfnamefont {P.~E.}\ \bibnamefont
  {Dargel}}, \bibinfo {author} {\bibfnamefont {A.}~\bibnamefont {W\"ollert}},
  \bibinfo {author} {\bibfnamefont {A.}~\bibnamefont {Honecker}}, \bibinfo
  {author} {\bibfnamefont {I.~P.}\ \bibnamefont {McCulloch}}, \bibinfo {author}
  {\bibfnamefont {U.}~\bibnamefont {Schollw\"ock}},\ and\ \bibinfo {author}
  {\bibfnamefont {T.}~\bibnamefont {Pruschke}},\ }\href
  {https://doi.org/10.1103/PhysRevB.85.205119} {\bibfield  {journal} {\bibinfo
  {journal} {Phys. Rev. B}\ }\textbf {\bibinfo {volume} {85}},\ \bibinfo
  {pages} {205119} (\bibinfo {year} {2012}{\natexlab{a}})}\BibitemShut
  {NoStop}%
\bibitem [{\citenamefont {K\"uhner}\ and\ \citenamefont
  {White}(1999)}]{White_cv_1999}%
  \BibitemOpen
  \bibfield  {author} {\bibinfo {author} {\bibfnamefont {T.~D.}\ \bibnamefont
  {K\"uhner}}\ and\ \bibinfo {author} {\bibfnamefont {S.~R.}\ \bibnamefont
  {White}},\ }\href {https://doi.org/10.1103/PhysRevB.60.335} {\bibfield
  {journal} {\bibinfo  {journal} {Phys. Rev. B}\ }\textbf {\bibinfo {volume}
  {60}},\ \bibinfo {pages} {335} (\bibinfo {year} {1999})}\BibitemShut
  {NoStop}%
\bibitem [{\citenamefont {Jeckelmann}(2002)}]{Jeckelmann_ddmrg_2002}%
  \BibitemOpen
  \bibfield  {author} {\bibinfo {author} {\bibfnamefont {E.}~\bibnamefont
  {Jeckelmann}},\ }\href {https://doi.org/10.1103/PhysRevB.66.045114}
  {\bibfield  {journal} {\bibinfo  {journal} {Phys. Rev. B}\ }\textbf {\bibinfo
  {volume} {66}},\ \bibinfo {pages} {045114} (\bibinfo {year}
  {2002})}\BibitemShut {NoStop}%
\bibitem [{\citenamefont {Jeckelmann}(2008)}]{Jeckelmann_ddmrg_2008}%
  \BibitemOpen
  \bibfield  {author} {\bibinfo {author} {\bibfnamefont {E.}~\bibnamefont
  {Jeckelmann}},\ }\href {https://doi.org/10.1143/PTPS.176.143} {\bibfield
  {journal} {\bibinfo  {journal} {Progress of Theoretical Physics Supplement}\
  }\textbf {\bibinfo {volume} {176}},\ \bibinfo {pages} {143} (\bibinfo {year}
  {2008})}\BibitemShut {NoStop}%
\bibitem [{\citenamefont {Weichselbaum}\ \emph {et~al.}(2009)\citenamefont
  {Weichselbaum}, \citenamefont {Verstraete}, \citenamefont {Schollw\"ock},
  \citenamefont {Cirac},\ and\ \citenamefont {von
  Delft}}]{Weichselbaum_unfiying_nrg_vmps_2009}%
  \BibitemOpen
  \bibfield  {author} {\bibinfo {author} {\bibfnamefont {A.}~\bibnamefont
  {Weichselbaum}}, \bibinfo {author} {\bibfnamefont {F.}~\bibnamefont
  {Verstraete}}, \bibinfo {author} {\bibfnamefont {U.}~\bibnamefont
  {Schollw\"ock}}, \bibinfo {author} {\bibfnamefont {J.~I.}\ \bibnamefont
  {Cirac}},\ and\ \bibinfo {author} {\bibfnamefont {J.}~\bibnamefont {von
  Delft}},\ }\href {https://doi.org/10.1103/PhysRevB.80.165117} {\bibfield
  {journal} {\bibinfo  {journal} {Phys. Rev. B}\ }\textbf {\bibinfo {volume}
  {80}},\ \bibinfo {pages} {165117} (\bibinfo {year} {2009})}\BibitemShut
  {NoStop}%
\bibitem [{\citenamefont {Paech}\ and\ \citenamefont
  {Jeckelmann}(2014)}]{Jeckelmann_deconvolution_blind_2014}%
  \BibitemOpen
  \bibfield  {author} {\bibinfo {author} {\bibfnamefont {M.}~\bibnamefont
  {Paech}}\ and\ \bibinfo {author} {\bibfnamefont {E.}~\bibnamefont
  {Jeckelmann}},\ }\href {https://doi.org/10.1103/PhysRevB.89.195101}
  {\bibfield  {journal} {\bibinfo  {journal} {Phys. Rev. B}\ }\textbf {\bibinfo
  {volume} {89}},\ \bibinfo {pages} {195101} (\bibinfo {year}
  {2014})}\BibitemShut {NoStop}%
\bibitem [{\citenamefont {Nocera}\ and\ \citenamefont
  {Alvarez}(2016)}]{Alvarez_cv_krylov_2016}%
  \BibitemOpen
  \bibfield  {author} {\bibinfo {author} {\bibfnamefont {A.}~\bibnamefont
  {Nocera}}\ and\ \bibinfo {author} {\bibfnamefont {G.}~\bibnamefont
  {Alvarez}},\ }\href {https://doi.org/10.1103/PhysRevE.94.053308} {\bibfield
  {journal} {\bibinfo  {journal} {Phys. Rev. E}\ }\textbf {\bibinfo {volume}
  {94}},\ \bibinfo {pages} {053308} (\bibinfo {year} {2016})}\BibitemShut
  {NoStop}%
\bibitem [{\citenamefont {Nocera}\ and\ \citenamefont
  {Alvarez}(2022)}]{Alvarez_cv_rootN_2022}%
  \BibitemOpen
  \bibfield  {author} {\bibinfo {author} {\bibfnamefont {A.}~\bibnamefont
  {Nocera}}\ and\ \bibinfo {author} {\bibfnamefont {G.}~\bibnamefont
  {Alvarez}},\ }\href {https://doi.org/10.1103/PhysRevB.106.205106} {\bibfield
  {journal} {\bibinfo  {journal} {Phys. Rev. B}\ }\textbf {\bibinfo {volume}
  {106}},\ \bibinfo {pages} {205106} (\bibinfo {year} {2022})}\BibitemShut
  {NoStop}%
\bibitem [{\citenamefont {Holzner}\ \emph {et~al.}(2011)\citenamefont
  {Holzner}, \citenamefont {Weichselbaum}, \citenamefont {McCulloch},
  \citenamefont {Schollw\"ock},\ and\ \citenamefont {von
  Delft}}]{Holzner_chebyshev_spectra_2011}%
  \BibitemOpen
  \bibfield  {author} {\bibinfo {author} {\bibfnamefont {A.}~\bibnamefont
  {Holzner}}, \bibinfo {author} {\bibfnamefont {A.}~\bibnamefont
  {Weichselbaum}}, \bibinfo {author} {\bibfnamefont {I.~P.}\ \bibnamefont
  {McCulloch}}, \bibinfo {author} {\bibfnamefont {U.}~\bibnamefont
  {Schollw\"ock}},\ and\ \bibinfo {author} {\bibfnamefont {J.}~\bibnamefont
  {von Delft}},\ }\href {https://doi.org/10.1103/PhysRevB.83.195115} {\bibfield
   {journal} {\bibinfo  {journal} {Phys. Rev. B}\ }\textbf {\bibinfo {volume}
  {83}},\ \bibinfo {pages} {195115} (\bibinfo {year} {2011})}\BibitemShut
  {NoStop}%
\bibitem [{\citenamefont {Wolf}\ \emph
  {et~al.}(2015{\natexlab{a}})\citenamefont {Wolf}, \citenamefont {Justiniano},
  \citenamefont {McCulloch},\ and\ \citenamefont
  {Schollw\"ock}}]{Wolf_chebyshev_time_2015}%
  \BibitemOpen
  \bibfield  {author} {\bibinfo {author} {\bibfnamefont {F.~A.}\ \bibnamefont
  {Wolf}}, \bibinfo {author} {\bibfnamefont {J.~A.}\ \bibnamefont
  {Justiniano}}, \bibinfo {author} {\bibfnamefont {I.~P.}\ \bibnamefont
  {McCulloch}},\ and\ \bibinfo {author} {\bibfnamefont {U.}~\bibnamefont
  {Schollw\"ock}},\ }\href {https://doi.org/10.1103/PhysRevB.91.115144}
  {\bibfield  {journal} {\bibinfo  {journal} {Phys. Rev. B}\ }\textbf {\bibinfo
  {volume} {91}},\ \bibinfo {pages} {115144} (\bibinfo {year}
  {2015}{\natexlab{a}})}\BibitemShut {NoStop}%
\bibitem [{\citenamefont {Halimeh}\ \emph {et~al.}(2015)\citenamefont
  {Halimeh}, \citenamefont {Kolley},\ and\ \citenamefont
  {McCulloch}}]{Halimeh_chebyshev_time_2015}%
  \BibitemOpen
  \bibfield  {author} {\bibinfo {author} {\bibfnamefont {J.~C.}\ \bibnamefont
  {Halimeh}}, \bibinfo {author} {\bibfnamefont {F.}~\bibnamefont {Kolley}},\
  and\ \bibinfo {author} {\bibfnamefont {I.~P.}\ \bibnamefont {McCulloch}},\
  }\href {https://doi.org/10.1103/PhysRevB.92.115130} {\bibfield  {journal}
  {\bibinfo  {journal} {Phys. Rev. B}\ }\textbf {\bibinfo {volume} {92}},\
  \bibinfo {pages} {115130} (\bibinfo {year} {2015})}\BibitemShut {NoStop}%
\bibitem [{\citenamefont {Xie}\ \emph {et~al.}(2018)\citenamefont {Xie},
  \citenamefont {Huang}, \citenamefont {Han}, \citenamefont {Yan},
  \citenamefont {Zhao}, \citenamefont {Xie}, \citenamefont {Liao},\ and\
  \citenamefont {Xiang}}]{Xiang_chebyshev_reorth_2018}%
  \BibitemOpen
  \bibfield  {author} {\bibinfo {author} {\bibfnamefont {H.~D.}\ \bibnamefont
  {Xie}}, \bibinfo {author} {\bibfnamefont {R.~Z.}\ \bibnamefont {Huang}},
  \bibinfo {author} {\bibfnamefont {X.~J.}\ \bibnamefont {Han}}, \bibinfo
  {author} {\bibfnamefont {X.}~\bibnamefont {Yan}}, \bibinfo {author}
  {\bibfnamefont {H.~H.}\ \bibnamefont {Zhao}}, \bibinfo {author}
  {\bibfnamefont {Z.~Y.}\ \bibnamefont {Xie}}, \bibinfo {author} {\bibfnamefont
  {H.~J.}\ \bibnamefont {Liao}},\ and\ \bibinfo {author} {\bibfnamefont
  {T.}~\bibnamefont {Xiang}},\ }\href
  {https://doi.org/10.1103/PhysRevB.97.075111} {\bibfield  {journal} {\bibinfo
  {journal} {Phys. Rev. B}\ }\textbf {\bibinfo {volume} {97}},\ \bibinfo
  {pages} {075111} (\bibinfo {year} {2018})}\BibitemShut {NoStop}%
\bibitem [{\citenamefont {Vidal}(2003)}]{Vidal_cononical_form_2003}%
  \BibitemOpen
  \bibfield  {author} {\bibinfo {author} {\bibfnamefont {G.}~\bibnamefont
  {Vidal}},\ }\href {https://doi.org/10.1103/PhysRevLett.91.147902} {\bibfield
  {journal} {\bibinfo  {journal} {Phys. Rev. Lett.}\ }\textbf {\bibinfo
  {volume} {91}},\ \bibinfo {pages} {147902} (\bibinfo {year}
  {2003})}\BibitemShut {NoStop}%
\bibitem [{\citenamefont {Vidal}(2004)}]{Vidal_time_evolution_2004}%
  \BibitemOpen
  \bibfield  {author} {\bibinfo {author} {\bibfnamefont {G.}~\bibnamefont
  {Vidal}},\ }\href {https://doi.org/10.1103/PhysRevLett.93.040502} {\bibfield
  {journal} {\bibinfo  {journal} {Phys. Rev. Lett.}\ }\textbf {\bibinfo
  {volume} {93}},\ \bibinfo {pages} {040502} (\bibinfo {year}
  {2004})}\BibitemShut {NoStop}%
\bibitem [{\citenamefont {White}\ and\ \citenamefont
  {Feiguin}(2004)}]{White_tDMRG_2004}%
  \BibitemOpen
  \bibfield  {author} {\bibinfo {author} {\bibfnamefont {S.~R.}\ \bibnamefont
  {White}}\ and\ \bibinfo {author} {\bibfnamefont {A.~E.}\ \bibnamefont
  {Feiguin}},\ }\href {https://doi.org/10.1103/PhysRevLett.93.076401}
  {\bibfield  {journal} {\bibinfo  {journal} {Phys. Rev. Lett.}\ }\textbf
  {\bibinfo {volume} {93}},\ \bibinfo {pages} {076401} (\bibinfo {year}
  {2004})}\BibitemShut {NoStop}%
\bibitem [{\citenamefont {Daley}\ \emph {et~al.}(2004)\citenamefont {Daley},
  \citenamefont {Kollath}, \citenamefont {Schollwöck},\ and\ \citenamefont
  {Vidal}}]{Daley_tebd_2004}%
  \BibitemOpen
  \bibfield  {author} {\bibinfo {author} {\bibfnamefont {A.~J.}\ \bibnamefont
  {Daley}}, \bibinfo {author} {\bibfnamefont {C.}~\bibnamefont {Kollath}},
  \bibinfo {author} {\bibfnamefont {U.}~\bibnamefont {Schollwöck}},\ and\
  \bibinfo {author} {\bibfnamefont {G.}~\bibnamefont {Vidal}},\ }\href
  {https://doi.org/10.1088/1742-5468/2004/04/P04005} {\bibfield  {journal}
  {\bibinfo  {journal} {Journal of Statistical Mechanics: Theory and
  Experiment}\ }\textbf {\bibinfo {volume} {2004}},\ \bibinfo {pages} {P04005}
  (\bibinfo {year} {2004})}\BibitemShut {NoStop}%
\bibitem [{\citenamefont {Verstraete}\ \emph {et~al.}(2004)\citenamefont
  {Verstraete}, \citenamefont {Garc\'{\i}a-Ripoll},\ and\ \citenamefont
  {Cirac}}]{Frank_time_finite_temp_2004}%
  \BibitemOpen
  \bibfield  {author} {\bibinfo {author} {\bibfnamefont {F.}~\bibnamefont
  {Verstraete}}, \bibinfo {author} {\bibfnamefont {J.~J.}\ \bibnamefont
  {Garc\'{\i}a-Ripoll}},\ and\ \bibinfo {author} {\bibfnamefont {J.~I.}\
  \bibnamefont {Cirac}},\ }\href
  {https://doi.org/10.1103/PhysRevLett.93.207204} {\bibfield  {journal}
  {\bibinfo  {journal} {Phys. Rev. Lett.}\ }\textbf {\bibinfo {volume} {93}},\
  \bibinfo {pages} {207204} (\bibinfo {year} {2004})}\BibitemShut {NoStop}%
\bibitem [{\citenamefont {Zwolak}\ and\ \citenamefont
  {Vidal}(2004)}]{Vidal_time_evolution_mixed_state_2004}%
  \BibitemOpen
  \bibfield  {author} {\bibinfo {author} {\bibfnamefont {M.}~\bibnamefont
  {Zwolak}}\ and\ \bibinfo {author} {\bibfnamefont {G.}~\bibnamefont {Vidal}},\
  }\href {https://doi.org/10.1103/PhysRevLett.93.207205} {\bibfield  {journal}
  {\bibinfo  {journal} {Phys. Rev. Lett.}\ }\textbf {\bibinfo {volume} {93}},\
  \bibinfo {pages} {207205} (\bibinfo {year} {2004})}\BibitemShut {NoStop}%
\bibitem [{\citenamefont {Zaletel}\ \emph {et~al.}(2015)\citenamefont
  {Zaletel}, \citenamefont {Mong}, \citenamefont {Karrasch}, \citenamefont
  {Moore},\ and\ \citenamefont {Pollmann}}]{Pollmann_MPO_time_evolution_2015}%
  \BibitemOpen
  \bibfield  {author} {\bibinfo {author} {\bibfnamefont {M.~P.}\ \bibnamefont
  {Zaletel}}, \bibinfo {author} {\bibfnamefont {R.~S.~K.}\ \bibnamefont
  {Mong}}, \bibinfo {author} {\bibfnamefont {C.}~\bibnamefont {Karrasch}},
  \bibinfo {author} {\bibfnamefont {J.~E.}\ \bibnamefont {Moore}},\ and\
  \bibinfo {author} {\bibfnamefont {F.}~\bibnamefont {Pollmann}},\ }\href
  {https://doi.org/10.1103/PhysRevB.91.165112} {\bibfield  {journal} {\bibinfo
  {journal} {Phys. Rev. B}\ }\textbf {\bibinfo {volume} {91}},\ \bibinfo
  {pages} {165112} (\bibinfo {year} {2015})}\BibitemShut {NoStop}%
\bibitem [{\citenamefont {García-Ripoll}(2006)}]{Juan_krylov_global_2006}%
  \BibitemOpen
  \bibfield  {author} {\bibinfo {author} {\bibfnamefont {J.~J.}\ \bibnamefont
  {García-Ripoll}},\ }\href {https://doi.org/10.1088/1367-2630/8/12/305}
  {\bibfield  {journal} {\bibinfo  {journal} {New Journal of Physics}\ }\textbf
  {\bibinfo {volume} {8}},\ \bibinfo {pages} {305} (\bibinfo {year}
  {2006})}\BibitemShut {NoStop}%
\bibitem [{\citenamefont {Dargel}\ \emph
  {et~al.}(2012{\natexlab{b}})\citenamefont {Dargel}, \citenamefont
  {W\"ollert}, \citenamefont {Honecker}, \citenamefont {McCulloch},
  \citenamefont {Schollw\"ock},\ and\ \citenamefont
  {Pruschke}}]{Dargel_krylov_global_2012}%
  \BibitemOpen
  \bibfield  {author} {\bibinfo {author} {\bibfnamefont {P.~E.}\ \bibnamefont
  {Dargel}}, \bibinfo {author} {\bibfnamefont {A.}~\bibnamefont {W\"ollert}},
  \bibinfo {author} {\bibfnamefont {A.}~\bibnamefont {Honecker}}, \bibinfo
  {author} {\bibfnamefont {I.~P.}\ \bibnamefont {McCulloch}}, \bibinfo {author}
  {\bibfnamefont {U.}~\bibnamefont {Schollw\"ock}},\ and\ \bibinfo {author}
  {\bibfnamefont {T.}~\bibnamefont {Pruschke}},\ }\href
  {https://doi.org/10.1103/PhysRevB.85.205119} {\bibfield  {journal} {\bibinfo
  {journal} {Phys. Rev. B}\ }\textbf {\bibinfo {volume} {85}},\ \bibinfo
  {pages} {205119} (\bibinfo {year} {2012}{\natexlab{b}})}\BibitemShut
  {NoStop}%
\bibitem [{\citenamefont {Wall}\ and\ \citenamefont
  {Carr}(2012)}]{Wall_krylov_global_2012}%
  \BibitemOpen
  \bibfield  {author} {\bibinfo {author} {\bibfnamefont {M.~L.}\ \bibnamefont
  {Wall}}\ and\ \bibinfo {author} {\bibfnamefont {L.~D.}\ \bibnamefont
  {Carr}},\ }\href {https://doi.org/10.1088/1367-2630/14/12/125015} {\bibfield
  {journal} {\bibinfo  {journal} {New Journal of Physics}\ }\textbf {\bibinfo
  {volume} {14}},\ \bibinfo {pages} {125015} (\bibinfo {year}
  {2012})}\BibitemShut {NoStop}%
\bibitem [{\citenamefont {Feiguin}\ and\ \citenamefont
  {White}(2005)}]{Feiguin_tdmrg_2005}%
  \BibitemOpen
  \bibfield  {author} {\bibinfo {author} {\bibfnamefont {A.~E.}\ \bibnamefont
  {Feiguin}}\ and\ \bibinfo {author} {\bibfnamefont {S.~R.}\ \bibnamefont
  {White}},\ }\href {https://doi.org/10.1103/PhysRevB.72.020404} {\bibfield
  {journal} {\bibinfo  {journal} {Phys. Rev. B}\ }\textbf {\bibinfo {volume}
  {72}},\ \bibinfo {pages} {020404} (\bibinfo {year} {2005})}\BibitemShut
  {NoStop}%
\bibitem [{\citenamefont {Schmitteckert}(2004)}]{Peter_krylov_local_2004}%
  \BibitemOpen
  \bibfield  {author} {\bibinfo {author} {\bibfnamefont {P.}~\bibnamefont
  {Schmitteckert}},\ }\href {https://doi.org/10.1103/PhysRevB.70.121302}
  {\bibfield  {journal} {\bibinfo  {journal} {Phys. Rev. B}\ }\textbf {\bibinfo
  {volume} {70}},\ \bibinfo {pages} {121302} (\bibinfo {year}
  {2004})}\BibitemShut {NoStop}%
\bibitem [{\citenamefont {Rodriguez}\ \emph {et~al.}(2006)\citenamefont
  {Rodriguez}, \citenamefont {Manmana}, \citenamefont {Rigol}, \citenamefont
  {Noack},\ and\ \citenamefont {Muramatsu}}]{Rodriguez_tdmrg_2006}%
  \BibitemOpen
  \bibfield  {author} {\bibinfo {author} {\bibfnamefont {K.}~\bibnamefont
  {Rodriguez}}, \bibinfo {author} {\bibfnamefont {S.~R.}\ \bibnamefont
  {Manmana}}, \bibinfo {author} {\bibfnamefont {M.}~\bibnamefont {Rigol}},
  \bibinfo {author} {\bibfnamefont {R.~M.}\ \bibnamefont {Noack}},\ and\
  \bibinfo {author} {\bibfnamefont {A.}~\bibnamefont {Muramatsu}},\ }\href
  {https://doi.org/10.1088/1367-2630/8/8/169} {\bibfield  {journal} {\bibinfo
  {journal} {New Journal of Physics}\ }\textbf {\bibinfo {volume} {8}},\
  \bibinfo {pages} {169} (\bibinfo {year} {2006})}\BibitemShut {NoStop}%
\bibitem [{\citenamefont {Ronca}\ \emph {et~al.}(2017)\citenamefont {Ronca},
  \citenamefont {Li}, \citenamefont {Jimenez-Hoyos},\ and\ \citenamefont
  {Chan}}]{Garnet_tdmrg_2017}%
  \BibitemOpen
  \bibfield  {author} {\bibinfo {author} {\bibfnamefont {E.}~\bibnamefont
  {Ronca}}, \bibinfo {author} {\bibfnamefont {Z.}~\bibnamefont {Li}}, \bibinfo
  {author} {\bibfnamefont {C.~A.}\ \bibnamefont {Jimenez-Hoyos}},\ and\
  \bibinfo {author} {\bibfnamefont {G.~K.-L.}\ \bibnamefont {Chan}},\ }\href
  {https://doi.org/10.1021/acs.jctc.7b00682} {\bibfield  {journal} {\bibinfo
  {journal} {Journal of Chemical Theory and Computation}\ }\textbf {\bibinfo
  {volume} {13}},\ \bibinfo {pages} {5560} (\bibinfo {year}
  {2017})}\BibitemShut {NoStop}%
\bibitem [{\citenamefont {Haegeman}\ \emph {et~al.}(2011)\citenamefont
  {Haegeman}, \citenamefont {Cirac}, \citenamefont {Osborne}, \citenamefont
  {Pi\ifmmode~\check{z}\else \v{z}\fi{}orn}, \citenamefont {Verschelde},\ and\
  \citenamefont {Verstraete}}]{tdvp_rk_2011}%
  \BibitemOpen
  \bibfield  {author} {\bibinfo {author} {\bibfnamefont {J.}~\bibnamefont
  {Haegeman}}, \bibinfo {author} {\bibfnamefont {J.~I.}\ \bibnamefont {Cirac}},
  \bibinfo {author} {\bibfnamefont {T.~J.}\ \bibnamefont {Osborne}}, \bibinfo
  {author} {\bibfnamefont {I.}~\bibnamefont {Pi\ifmmode~\check{z}\else
  \v{z}\fi{}orn}}, \bibinfo {author} {\bibfnamefont {H.}~\bibnamefont
  {Verschelde}},\ and\ \bibinfo {author} {\bibfnamefont {F.}~\bibnamefont
  {Verstraete}},\ }\href {https://doi.org/10.1103/PhysRevLett.107.070601}
  {\bibfield  {journal} {\bibinfo  {journal} {Phys. Rev. Lett.}\ }\textbf
  {\bibinfo {volume} {107}},\ \bibinfo {pages} {070601} (\bibinfo {year}
  {2011})}\BibitemShut {NoStop}%
\bibitem [{\citenamefont {Haegeman}\ \emph {et~al.}(2013)\citenamefont
  {Haegeman}, \citenamefont {Osborne},\ and\ \citenamefont
  {Verstraete}}]{tdvp_tangent_2013}%
  \BibitemOpen
  \bibfield  {author} {\bibinfo {author} {\bibfnamefont {J.}~\bibnamefont
  {Haegeman}}, \bibinfo {author} {\bibfnamefont {T.~J.}\ \bibnamefont
  {Osborne}},\ and\ \bibinfo {author} {\bibfnamefont {F.}~\bibnamefont
  {Verstraete}},\ }\href {https://doi.org/10.1103/PhysRevB.88.075133}
  {\bibfield  {journal} {\bibinfo  {journal} {Phys. Rev. B}\ }\textbf {\bibinfo
  {volume} {88}},\ \bibinfo {pages} {075133} (\bibinfo {year}
  {2013})}\BibitemShut {NoStop}%
\bibitem [{\citenamefont {Haegeman}\ \emph {et~al.}(2016)\citenamefont
  {Haegeman}, \citenamefont {Lubich}, \citenamefont {Oseledets}, \citenamefont
  {Vandereycken},\ and\ \citenamefont {Verstraete}}]{tdvp_local_2016}%
  \BibitemOpen
  \bibfield  {author} {\bibinfo {author} {\bibfnamefont {J.}~\bibnamefont
  {Haegeman}}, \bibinfo {author} {\bibfnamefont {C.}~\bibnamefont {Lubich}},
  \bibinfo {author} {\bibfnamefont {I.}~\bibnamefont {Oseledets}}, \bibinfo
  {author} {\bibfnamefont {B.}~\bibnamefont {Vandereycken}},\ and\ \bibinfo
  {author} {\bibfnamefont {F.}~\bibnamefont {Verstraete}},\ }\href
  {https://doi.org/10.1103/PhysRevB.94.165116} {\bibfield  {journal} {\bibinfo
  {journal} {Phys. Rev. B}\ }\textbf {\bibinfo {volume} {94}},\ \bibinfo
  {pages} {165116} (\bibinfo {year} {2016})}\BibitemShut {NoStop}%
\bibitem [{\citenamefont {Vanderstraeten}\ \emph {et~al.}(2019)\citenamefont
  {Vanderstraeten}, \citenamefont {Haegeman},\ and\ \citenamefont
  {Verstraete}}]{tdvp_tangent_uMPS_2019}%
  \BibitemOpen
  \bibfield  {author} {\bibinfo {author} {\bibfnamefont {L.}~\bibnamefont
  {Vanderstraeten}}, \bibinfo {author} {\bibfnamefont {J.}~\bibnamefont
  {Haegeman}},\ and\ \bibinfo {author} {\bibfnamefont {F.}~\bibnamefont
  {Verstraete}},\ }\href {https://doi.org/10.21468/SciPostPhysLectNotes.7}
  {\bibfield  {journal} {\bibinfo  {journal} {SciPost Phys. Lect. Notes}\ ,\
  \bibinfo {pages} {7}} (\bibinfo {year} {2019})}\BibitemShut {NoStop}%
\bibitem [{\citenamefont {Paeckel}\ \emph {et~al.}(2019)\citenamefont
  {Paeckel}, \citenamefont {Köhler}, \citenamefont {Swoboda}, \citenamefont
  {Manmana}, \citenamefont {Schollwöck},\ and\ \citenamefont
  {Hubig}}]{Paeckel_time_evolution_review_2019}%
  \BibitemOpen
  \bibfield  {author} {\bibinfo {author} {\bibfnamefont {S.}~\bibnamefont
  {Paeckel}}, \bibinfo {author} {\bibfnamefont {T.}~\bibnamefont {Köhler}},
  \bibinfo {author} {\bibfnamefont {A.}~\bibnamefont {Swoboda}}, \bibinfo
  {author} {\bibfnamefont {S.~R.}\ \bibnamefont {Manmana}}, \bibinfo {author}
  {\bibfnamefont {U.}~\bibnamefont {Schollwöck}},\ and\ \bibinfo {author}
  {\bibfnamefont {C.}~\bibnamefont {Hubig}},\ }\href
  {https://doi.org/https://doi.org/10.1016/j.aop.2019.167998} {\bibfield
  {journal} {\bibinfo  {journal} {Annals of Physics}\ }\textbf {\bibinfo
  {volume} {411}},\ \bibinfo {pages} {167998} (\bibinfo {year}
  {2019})}\BibitemShut {NoStop}%
\bibitem [{\citenamefont {Calabrese}\ and\ \citenamefont
  {Cardy}(2005)}]{Calabrese_entangle_2005}%
  \BibitemOpen
  \bibfield  {author} {\bibinfo {author} {\bibfnamefont {P.}~\bibnamefont
  {Calabrese}}\ and\ \bibinfo {author} {\bibfnamefont {J.}~\bibnamefont
  {Cardy}},\ }\href {https://doi.org/10.1088/1742-5468/2005/04/P04010}
  {\bibfield  {journal} {\bibinfo  {journal} {Journal of Statistical Mechanics:
  Theory and Experiment}\ }\textbf {\bibinfo {volume} {2005}},\ \bibinfo
  {pages} {P04010} (\bibinfo {year} {2005})}\BibitemShut {NoStop}%
\bibitem [{\citenamefont {Polkovnikov}\ \emph {et~al.}(2011)\citenamefont
  {Polkovnikov}, \citenamefont {Sengupta}, \citenamefont {Silva},\ and\
  \citenamefont {Vengalattore}}]{Polkovnikov_2011}%
  \BibitemOpen
  \bibfield  {author} {\bibinfo {author} {\bibfnamefont {A.}~\bibnamefont
  {Polkovnikov}}, \bibinfo {author} {\bibfnamefont {K.}~\bibnamefont
  {Sengupta}}, \bibinfo {author} {\bibfnamefont {A.}~\bibnamefont {Silva}},\
  and\ \bibinfo {author} {\bibfnamefont {M.}~\bibnamefont {Vengalattore}},\
  }\href {https://doi.org/10.1103/RevModPhys.83.863} {\bibfield  {journal}
  {\bibinfo  {journal} {Rev. Mod. Phys.}\ }\textbf {\bibinfo {volume} {83}},\
  \bibinfo {pages} {863} (\bibinfo {year} {2011})}\BibitemShut {NoStop}%
\bibitem [{\citenamefont {Eisert}\ and\ \citenamefont
  {Osborne}(2006)}]{Tobias_entanglement_time_2006}%
  \BibitemOpen
  \bibfield  {author} {\bibinfo {author} {\bibfnamefont {J.}~\bibnamefont
  {Eisert}}\ and\ \bibinfo {author} {\bibfnamefont {T.~J.}\ \bibnamefont
  {Osborne}},\ }\href {https://doi.org/10.1103/PhysRevLett.97.150404}
  {\bibfield  {journal} {\bibinfo  {journal} {Phys. Rev. Lett.}\ }\textbf
  {\bibinfo {volume} {97}},\ \bibinfo {pages} {150404} (\bibinfo {year}
  {2006})}\BibitemShut {NoStop}%
\bibitem [{\citenamefont
  {Osborne}(2006)}]{Tobias_entanglement_time_appro_2006}%
  \BibitemOpen
  \bibfield  {author} {\bibinfo {author} {\bibfnamefont {T.~J.}\ \bibnamefont
  {Osborne}},\ }\href {https://doi.org/10.1103/PhysRevLett.97.157202}
  {\bibfield  {journal} {\bibinfo  {journal} {Phys. Rev. Lett.}\ }\textbf
  {\bibinfo {volume} {97}},\ \bibinfo {pages} {157202} (\bibinfo {year}
  {2006})}\BibitemShut {NoStop}%
\bibitem [{\citenamefont
  {Bravyi}(2007)}]{Bravyi_upper_bounds_unitary_evolution_2007}%
  \BibitemOpen
  \bibfield  {author} {\bibinfo {author} {\bibfnamefont {S.}~\bibnamefont
  {Bravyi}},\ }\href {https://doi.org/10.1103/PhysRevA.76.052319} {\bibfield
  {journal} {\bibinfo  {journal} {Phys. Rev. A}\ }\textbf {\bibinfo {volume}
  {76}},\ \bibinfo {pages} {052319} (\bibinfo {year} {2007})}\BibitemShut
  {NoStop}%
\bibitem [{\citenamefont {Mari{\"e}n}\ \emph {et~al.}(2016)\citenamefont
  {Mari{\"e}n}, \citenamefont {Audenaert}, \citenamefont {Van~Acoleyen},\ and\
  \citenamefont {Verstraete}}]{Frank_upper_bounds_stability_2016}%
  \BibitemOpen
  \bibfield  {author} {\bibinfo {author} {\bibfnamefont {M.}~\bibnamefont
  {Mari{\"e}n}}, \bibinfo {author} {\bibfnamefont {K.~M.}\ \bibnamefont
  {Audenaert}}, \bibinfo {author} {\bibfnamefont {K.}~\bibnamefont
  {Van~Acoleyen}},\ and\ \bibinfo {author} {\bibfnamefont {F.}~\bibnamefont
  {Verstraete}},\ }\href@noop {} {\bibfield  {journal} {\bibinfo  {journal}
  {Communications in Mathematical Physics}\ }\textbf {\bibinfo {volume}
  {346}},\ \bibinfo {pages} {35} (\bibinfo {year} {2016})}\BibitemShut
  {NoStop}%
\bibitem [{\citenamefont {Alhambra}\ and\ \citenamefont
  {Cirac}(2021)}]{Cirac_tns_local_time_evolution_2021}%
  \BibitemOpen
  \bibfield  {author} {\bibinfo {author} {\bibfnamefont {A.~M.}\ \bibnamefont
  {Alhambra}}\ and\ \bibinfo {author} {\bibfnamefont {J.~I.}\ \bibnamefont
  {Cirac}},\ }\href {https://doi.org/10.1103/PRXQuantum.2.040331} {\bibfield
  {journal} {\bibinfo  {journal} {PRX Quantum}\ }\textbf {\bibinfo {volume}
  {2}},\ \bibinfo {pages} {040331} (\bibinfo {year} {2021})}\BibitemShut
  {NoStop}%
\bibitem [{\citenamefont {Kuwahara}\ \emph {et~al.}(2021)\citenamefont
  {Kuwahara}, \citenamefont {Alhambra},\ and\ \citenamefont
  {Anshu}}]{Kuwahara_improved_area_law_2021}%
  \BibitemOpen
  \bibfield  {author} {\bibinfo {author} {\bibfnamefont {T.}~\bibnamefont
  {Kuwahara}}, \bibinfo {author} {\bibfnamefont {A.~M.}\ \bibnamefont
  {Alhambra}},\ and\ \bibinfo {author} {\bibfnamefont {A.}~\bibnamefont
  {Anshu}},\ }\href {https://doi.org/10.1103/PhysRevX.11.011047} {\bibfield
  {journal} {\bibinfo  {journal} {Phys. Rev. X}\ }\textbf {\bibinfo {volume}
  {11}},\ \bibinfo {pages} {011047} (\bibinfo {year} {2021})}\BibitemShut
  {NoStop}%
\bibitem [{\citenamefont {White}\ and\ \citenamefont
  {Affleck}(2008)}]{White_linear_prediction_2008}%
  \BibitemOpen
  \bibfield  {author} {\bibinfo {author} {\bibfnamefont {S.~R.}\ \bibnamefont
  {White}}\ and\ \bibinfo {author} {\bibfnamefont {I.}~\bibnamefont
  {Affleck}},\ }\href {https://doi.org/10.1103/PhysRevB.77.134437} {\bibfield
  {journal} {\bibinfo  {journal} {Phys. Rev. B}\ }\textbf {\bibinfo {volume}
  {77}},\ \bibinfo {pages} {134437} (\bibinfo {year} {2008})}\BibitemShut
  {NoStop}%
\bibitem [{\citenamefont {Tian}\ and\ \citenamefont
  {White}(2021)}]{White_recursion_2021}%
  \BibitemOpen
  \bibfield  {author} {\bibinfo {author} {\bibfnamefont {Y.}~\bibnamefont
  {Tian}}\ and\ \bibinfo {author} {\bibfnamefont {S.~R.}\ \bibnamefont
  {White}},\ }\href {https://doi.org/10.1103/PhysRevB.103.125142} {\bibfield
  {journal} {\bibinfo  {journal} {Phys. Rev. B}\ }\textbf {\bibinfo {volume}
  {103}},\ \bibinfo {pages} {125142} (\bibinfo {year} {2021})}\BibitemShut
  {NoStop}%
\bibitem [{\citenamefont {Grundner}\ \emph {et~al.}(2024)\citenamefont
  {Grundner}, \citenamefont {Westhoff}, \citenamefont {Kugler}, \citenamefont
  {Parcollet},\ and\ \citenamefont {Schollw\"ock}}]{grundner2023complex}%
  \BibitemOpen
  \bibfield  {author} {\bibinfo {author} {\bibfnamefont {M.}~\bibnamefont
  {Grundner}}, \bibinfo {author} {\bibfnamefont {P.}~\bibnamefont {Westhoff}},
  \bibinfo {author} {\bibfnamefont {F.~B.}\ \bibnamefont {Kugler}}, \bibinfo
  {author} {\bibfnamefont {O.}~\bibnamefont {Parcollet}},\ and\ \bibinfo
  {author} {\bibfnamefont {U.}~\bibnamefont {Schollw\"ock}},\ }\href
  {https://doi.org/10.1103/PhysRevB.109.155124} {\bibfield  {journal} {\bibinfo
   {journal} {Phys. Rev. B}\ }\textbf {\bibinfo {volume} {109}},\ \bibinfo
  {pages} {155124} (\bibinfo {year} {2024})}\BibitemShut {NoStop}%
\bibitem [{Note1()}]{Note1}%
  \BibitemOpen
  \bibinfo {note} {In general, we want to use $\protect \hat {H}_{} - E_0$,
  where $E_0$ is the ground state energy for a gapless system considered in
  this work, but we may want to use a different energy for a gapped system to
  ensure a better convergence.}\BibitemShut {Stop}%
\bibitem [{\citenamefont {Wolf}\ \emph
  {et~al.}(2015{\natexlab{b}})\citenamefont {Wolf}, \citenamefont {Go},
  \citenamefont {McCulloch}, \citenamefont {Millis},\ and\ \citenamefont
  {Schollw\"ock}}]{Wolf_imaginary_time_2015}%
  \BibitemOpen
  \bibfield  {author} {\bibinfo {author} {\bibfnamefont {F.~A.}\ \bibnamefont
  {Wolf}}, \bibinfo {author} {\bibfnamefont {A.}~\bibnamefont {Go}}, \bibinfo
  {author} {\bibfnamefont {I.~P.}\ \bibnamefont {McCulloch}}, \bibinfo {author}
  {\bibfnamefont {A.~J.}\ \bibnamefont {Millis}},\ and\ \bibinfo {author}
  {\bibfnamefont {U.}~\bibnamefont {Schollw\"ock}},\ }\href
  {https://doi.org/10.1103/PhysRevX.5.041032} {\bibfield  {journal} {\bibinfo
  {journal} {Phys. Rev. X}\ }\textbf {\bibinfo {volume} {5}},\ \bibinfo {pages}
  {041032} (\bibinfo {year} {2015}{\natexlab{b}})}\BibitemShut {NoStop}%
\bibitem [{\citenamefont {Guther}\ \emph {et~al.}(2018)\citenamefont {Guther},
  \citenamefont {Dobrautz}, \citenamefont {Gunnarsson},\ and\ \citenamefont
  {Alavi}}]{Kai_complex_time_fciqmc_2018}%
  \BibitemOpen
  \bibfield  {author} {\bibinfo {author} {\bibfnamefont {K.}~\bibnamefont
  {Guther}}, \bibinfo {author} {\bibfnamefont {W.}~\bibnamefont {Dobrautz}},
  \bibinfo {author} {\bibfnamefont {O.}~\bibnamefont {Gunnarsson}},\ and\
  \bibinfo {author} {\bibfnamefont {A.}~\bibnamefont {Alavi}},\ }\href
  {https://doi.org/10.1103/PhysRevLett.121.056401} {\bibfield  {journal}
  {\bibinfo  {journal} {Phys. Rev. Lett.}\ }\textbf {\bibinfo {volume} {121}},\
  \bibinfo {pages} {056401} (\bibinfo {year} {2018})}\BibitemShut {NoStop}%
\bibitem [{\citenamefont {Lu}\ \emph {et~al.}(2014)\citenamefont {Lu},
  \citenamefont {H\"oppner}, \citenamefont {Gunnarsson},\ and\ \citenamefont
  {Haverkort}}]{Lu2014}%
  \BibitemOpen
  \bibfield  {author} {\bibinfo {author} {\bibfnamefont {Y.}~\bibnamefont
  {Lu}}, \bibinfo {author} {\bibfnamefont {M.}~\bibnamefont {H\"oppner}},
  \bibinfo {author} {\bibfnamefont {O.}~\bibnamefont {Gunnarsson}},\ and\
  \bibinfo {author} {\bibfnamefont {M.~W.}\ \bibnamefont {Haverkort}},\ }\href
  {https://doi.org/10.1103/PhysRevB.90.085102} {\bibfield  {journal} {\bibinfo
  {journal} {Phys. Rev. B}\ }\textbf {\bibinfo {volume} {90}},\ \bibinfo
  {pages} {085102} (\bibinfo {year} {2014})}\BibitemShut {NoStop}%
\bibitem [{\citenamefont {Lu}\ \emph {et~al.}(2019)\citenamefont {Lu},
  \citenamefont {Cao}, \citenamefont {Hansmann},\ and\ \citenamefont
  {Haverkort}}]{Lu2019}%
  \BibitemOpen
  \bibfield  {author} {\bibinfo {author} {\bibfnamefont {Y.}~\bibnamefont
  {Lu}}, \bibinfo {author} {\bibfnamefont {X.}~\bibnamefont {Cao}}, \bibinfo
  {author} {\bibfnamefont {P.}~\bibnamefont {Hansmann}},\ and\ \bibinfo
  {author} {\bibfnamefont {M.~W.}\ \bibnamefont {Haverkort}},\ }\href
  {https://doi.org/10.1103/PhysRevB.100.115134} {\bibfield  {journal} {\bibinfo
   {journal} {Phys. Rev. B}\ }\textbf {\bibinfo {volume} {100}},\ \bibinfo
  {pages} {115134} (\bibinfo {year} {2019})}\BibitemShut {NoStop}%
\bibitem [{\citenamefont {Cao}\ \emph {et~al.}(2021)\citenamefont {Cao},
  \citenamefont {Lu}, \citenamefont {Hansmann},\ and\ \citenamefont
  {Haverkort}}]{cao_tree_imp_2021}%
  \BibitemOpen
  \bibfield  {author} {\bibinfo {author} {\bibfnamefont {X.}~\bibnamefont
  {Cao}}, \bibinfo {author} {\bibfnamefont {Y.}~\bibnamefont {Lu}}, \bibinfo
  {author} {\bibfnamefont {P.}~\bibnamefont {Hansmann}},\ and\ \bibinfo
  {author} {\bibfnamefont {M.~W.}\ \bibnamefont {Haverkort}},\ }\href
  {https://doi.org/10.1103/PhysRevB.104.115119} {\bibfield  {journal} {\bibinfo
   {journal} {Phys. Rev. B}\ }\textbf {\bibinfo {volume} {104}},\ \bibinfo
  {pages} {115119} (\bibinfo {year} {2021})}\BibitemShut {NoStop}%
\bibitem [{\citenamefont {Stoudenmire}\ and\ \citenamefont
  {White}(2010)}]{Stoudenmire_metts_2010}%
  \BibitemOpen
  \bibfield  {author} {\bibinfo {author} {\bibfnamefont {E.~M.}\ \bibnamefont
  {Stoudenmire}}\ and\ \bibinfo {author} {\bibfnamefont {S.~R.}\ \bibnamefont
  {White}},\ }\href {https://doi.org/10.1088/1367-2630/12/5/055026} {\bibfield
  {journal} {\bibinfo  {journal} {New Journal of Physics}\ }\textbf {\bibinfo
  {volume} {12}},\ \bibinfo {pages} {055026} (\bibinfo {year}
  {2010})}\BibitemShut {NoStop}%
\bibitem [{\citenamefont {Weichselbaum}\ and\ \citenamefont {von
  Delft}(2007)}]{nrg_fabian_1}%
  \BibitemOpen
  \bibfield  {author} {\bibinfo {author} {\bibfnamefont {A.}~\bibnamefont
  {Weichselbaum}}\ and\ \bibinfo {author} {\bibfnamefont {J.}~\bibnamefont {von
  Delft}},\ }\href {https://doi.org/10.1103/PhysRevLett.99.076402} {\bibfield
  {journal} {\bibinfo  {journal} {Phys. Rev. Lett.}\ }\textbf {\bibinfo
  {volume} {99}},\ \bibinfo {pages} {076402} (\bibinfo {year}
  {2007})}\BibitemShut {NoStop}%
\bibitem [{\citenamefont {Peters}\ \emph {et~al.}(2006)\citenamefont {Peters},
  \citenamefont {Pruschke},\ and\ \citenamefont {Anders}}]{nrg_fabian_2}%
  \BibitemOpen
  \bibfield  {author} {\bibinfo {author} {\bibfnamefont {R.}~\bibnamefont
  {Peters}}, \bibinfo {author} {\bibfnamefont {T.}~\bibnamefont {Pruschke}},\
  and\ \bibinfo {author} {\bibfnamefont {F.~B.}\ \bibnamefont {Anders}},\
  }\href {https://doi.org/10.1103/PhysRevB.74.245114} {\bibfield  {journal}
  {\bibinfo  {journal} {Phys. Rev. B}\ }\textbf {\bibinfo {volume} {74}},\
  \bibinfo {pages} {245114} (\bibinfo {year} {2006})}\BibitemShut {NoStop}%
\bibitem [{\citenamefont {\ifmmode~\check{Z}\else \v{Z}\fi{}itko}\ and\
  \citenamefont {Pruschke}(2009)}]{nrg_fabian_3}%
  \BibitemOpen
  \bibfield  {author} {\bibinfo {author} {\bibfnamefont {R.}~\bibnamefont
  {\ifmmode~\check{Z}\else \v{Z}\fi{}itko}}\ and\ \bibinfo {author}
  {\bibfnamefont {T.}~\bibnamefont {Pruschke}},\ }\href
  {https://doi.org/10.1103/PhysRevB.79.085106} {\bibfield  {journal} {\bibinfo
  {journal} {Phys. Rev. B}\ }\textbf {\bibinfo {volume} {79}},\ \bibinfo
  {pages} {085106} (\bibinfo {year} {2009})}\BibitemShut {NoStop}%
\bibitem [{\citenamefont {Weichselbaum}(2012{\natexlab{a}})}]{nrg_fabian_4}%
  \BibitemOpen
  \bibfield  {author} {\bibinfo {author} {\bibfnamefont {A.}~\bibnamefont
  {Weichselbaum}},\ }\href {https://doi.org/10.1103/PhysRevB.86.245124}
  {\bibfield  {journal} {\bibinfo  {journal} {Phys. Rev. B}\ }\textbf {\bibinfo
  {volume} {86}},\ \bibinfo {pages} {245124} (\bibinfo {year}
  {2012}{\natexlab{a}})}\BibitemShut {NoStop}%
\bibitem [{\citenamefont {Lee}\ \emph {et~al.}(2017)\citenamefont {Lee},
  \citenamefont {von Delft},\ and\ \citenamefont
  {Weichselbaum}}]{nrg_fabian_5}%
  \BibitemOpen
  \bibfield  {author} {\bibinfo {author} {\bibfnamefont {S.-S.~B.}\
  \bibnamefont {Lee}}, \bibinfo {author} {\bibfnamefont {J.}~\bibnamefont {von
  Delft}},\ and\ \bibinfo {author} {\bibfnamefont {A.}~\bibnamefont
  {Weichselbaum}},\ }\href {https://doi.org/10.1103/PhysRevLett.119.236402}
  {\bibfield  {journal} {\bibinfo  {journal} {Phys. Rev. Lett.}\ }\textbf
  {\bibinfo {volume} {119}},\ \bibinfo {pages} {236402} (\bibinfo {year}
  {2017})}\BibitemShut {NoStop}%
\bibitem [{\citenamefont {Lee}\ and\ \citenamefont
  {Weichselbaum}(2016)}]{nrg_fabian_6}%
  \BibitemOpen
  \bibfield  {author} {\bibinfo {author} {\bibfnamefont {S.-S.~B.}\
  \bibnamefont {Lee}}\ and\ \bibinfo {author} {\bibfnamefont {A.}~\bibnamefont
  {Weichselbaum}},\ }\href {https://doi.org/10.1103/PhysRevB.94.235127}
  {\bibfield  {journal} {\bibinfo  {journal} {Phys. Rev. B}\ }\textbf {\bibinfo
  {volume} {94}},\ \bibinfo {pages} {235127} (\bibinfo {year}
  {2016})}\BibitemShut {NoStop}%
\bibitem [{\citenamefont {Weichselbaum}(2012{\natexlab{b}})}]{nrg_fabian_7}%
  \BibitemOpen
  \bibfield  {author} {\bibinfo {author} {\bibfnamefont {A.}~\bibnamefont
  {Weichselbaum}},\ }\href
  {https://doi.org/https://doi.org/10.1016/j.aop.2012.07.009} {\bibfield
  {journal} {\bibinfo  {journal} {Annals of Physics}\ }\textbf {\bibinfo
  {volume} {327}},\ \bibinfo {pages} {2972} (\bibinfo {year}
  {2012}{\natexlab{b}})}\BibitemShut {NoStop}%
\bibitem [{\citenamefont {Kugler}(2022)}]{fabian2022}%
  \BibitemOpen
  \bibfield  {author} {\bibinfo {author} {\bibfnamefont {F.~B.}\ \bibnamefont
  {Kugler}},\ }\href {https://doi.org/10.1103/PhysRevB.105.245132} {\bibfield
  {journal} {\bibinfo  {journal} {Phys. Rev. B}\ }\textbf {\bibinfo {volume}
  {105}},\ \bibinfo {pages} {245132} (\bibinfo {year} {2022})}\BibitemShut
  {NoStop}%
\bibitem [{\citenamefont {Luttinger}(1960)}]{Luttinger_1960}%
  \BibitemOpen
  \bibfield  {author} {\bibinfo {author} {\bibfnamefont {J.~M.}\ \bibnamefont
  {Luttinger}},\ }\href {https://doi.org/10.1103/PhysRev.119.1153} {\bibfield
  {journal} {\bibinfo  {journal} {Phys. Rev.}\ }\textbf {\bibinfo {volume}
  {119}},\ \bibinfo {pages} {1153} (\bibinfo {year} {1960})}\BibitemShut
  {NoStop}%
\bibitem [{\citenamefont {Luttinger}(1961)}]{Luttinger_1961}%
  \BibitemOpen
  \bibfield  {author} {\bibinfo {author} {\bibfnamefont {J.~M.}\ \bibnamefont
  {Luttinger}},\ }\href {https://doi.org/10.1103/PhysRev.121.942} {\bibfield
  {journal} {\bibinfo  {journal} {Phys. Rev.}\ }\textbf {\bibinfo {volume}
  {121}},\ \bibinfo {pages} {942} (\bibinfo {year} {1961})}\BibitemShut
  {NoStop}%
\bibitem [{\citenamefont {Raas}\ and\ \citenamefont
  {Uhrig}(2005)}]{Raas_siam_kondo_2005}%
  \BibitemOpen
  \bibfield  {author} {\bibinfo {author} {\bibfnamefont {C.}~\bibnamefont
  {Raas}}\ and\ \bibinfo {author} {\bibfnamefont {G.~S.}\ \bibnamefont
  {Uhrig}},\ }\href@noop {} {\bibfield  {journal} {\bibinfo  {journal} {The
  European Physical Journal B-Condensed Matter and Complex Systems}\ }\textbf
  {\bibinfo {volume} {45}},\ \bibinfo {pages} {293} (\bibinfo {year}
  {2005})}\BibitemShut {NoStop}%
\bibitem [{\citenamefont {Hewson}(1997)}]{hewson1997kondo}%
  \BibitemOpen
  \bibfield  {author} {\bibinfo {author} {\bibfnamefont {A.~C.}\ \bibnamefont
  {Hewson}},\ }\href@noop {} {\emph {\bibinfo {title} {The Kondo Problem to
  Heavy Fermions}}},\ \bibinfo {number} {2}\ (\bibinfo  {publisher} {Cambridge
  university press},\ \bibinfo {year} {1997})\BibitemShut {NoStop}%
\bibitem [{\citenamefont {Bauernfeind}\ \emph {et~al.}(2017)\citenamefont
  {Bauernfeind}, \citenamefont {Zingl}, \citenamefont {Triebl}, \citenamefont
  {Aichhorn},\ and\ \citenamefont {Evertz}}]{Daniel_fork_2017}%
  \BibitemOpen
  \bibfield  {author} {\bibinfo {author} {\bibfnamefont {D.}~\bibnamefont
  {Bauernfeind}}, \bibinfo {author} {\bibfnamefont {M.}~\bibnamefont {Zingl}},
  \bibinfo {author} {\bibfnamefont {R.}~\bibnamefont {Triebl}}, \bibinfo
  {author} {\bibfnamefont {M.}~\bibnamefont {Aichhorn}},\ and\ \bibinfo
  {author} {\bibfnamefont {H.~G.}\ \bibnamefont {Evertz}},\ }\href
  {https://doi.org/10.1103/PhysRevX.7.031013} {\bibfield  {journal} {\bibinfo
  {journal} {Phys. Rev. X}\ }\textbf {\bibinfo {volume} {7}},\ \bibinfo {pages}
  {031013} (\bibinfo {year} {2017})}\BibitemShut {NoStop}%
\bibitem [{\citenamefont {Kloss}\ \emph {et~al.}(2020)\citenamefont {Kloss},
  \citenamefont {Reichman},\ and\ \citenamefont {Lev}}]{Kloss_tree_2020}%
  \BibitemOpen
  \bibfield  {author} {\bibinfo {author} {\bibfnamefont {B.}~\bibnamefont
  {Kloss}}, \bibinfo {author} {\bibfnamefont {D.~R.}\ \bibnamefont
  {Reichman}},\ and\ \bibinfo {author} {\bibfnamefont {Y.~B.}\ \bibnamefont
  {Lev}},\ }\href {https://doi.org/10.21468/SciPostPhys.9.5.070} {\bibfield
  {journal} {\bibinfo  {journal} {SciPost Phys.}\ }\textbf {\bibinfo {volume}
  {9}},\ \bibinfo {pages} {070} (\bibinfo {year} {2020})}\BibitemShut {NoStop}%
\bibitem [{\citenamefont {White}\ \emph {et~al.}(2018)\citenamefont {White},
  \citenamefont {Zaletel}, \citenamefont {Mong},\ and\ \citenamefont
  {Refael}}]{White18}%
  \BibitemOpen
  \bibfield  {author} {\bibinfo {author} {\bibfnamefont {C.~D.}\ \bibnamefont
  {White}}, \bibinfo {author} {\bibfnamefont {M.}~\bibnamefont {Zaletel}},
  \bibinfo {author} {\bibfnamefont {R.~S.~K.}\ \bibnamefont {Mong}},\ and\
  \bibinfo {author} {\bibfnamefont {G.}~\bibnamefont {Refael}},\ }\href
  {https://doi.org/10.1103/PhysRevB.97.035127} {\bibfield  {journal} {\bibinfo
  {journal} {Phys. Rev. B}\ }\textbf {\bibinfo {volume} {97}},\ \bibinfo
  {pages} {035127} (\bibinfo {year} {2018})}\BibitemShut {NoStop}%
\bibitem [{\citenamefont {Ye}\ and\ \citenamefont {Chan}(2021)}]{Ye2021}%
  \BibitemOpen
  \bibfield  {author} {\bibinfo {author} {\bibfnamefont {E.}~\bibnamefont
  {Ye}}\ and\ \bibinfo {author} {\bibfnamefont {G.~K.-L.}\ \bibnamefont
  {Chan}},\ }\href {https://doi.org/10.1063/5.0047260} {\bibfield  {journal}
  {\bibinfo  {journal} {The Journal of Chemical Physics}\ }\textbf {\bibinfo
  {volume} {155}},\ \bibinfo {pages} {044104} (\bibinfo {year}
  {2021})}\BibitemShut {NoStop}%
\bibitem [{\citenamefont {Fr{\'\i}as-P{\'e}rez}\ \emph
  {et~al.}(2023)\citenamefont {Fr{\'\i}as-P{\'e}rez}, \citenamefont
  {Tagliacozzo},\ and\ \citenamefont {Ba{\~n}uls}}]{FriasPerez}%
  \BibitemOpen
  \bibfield  {author} {\bibinfo {author} {\bibfnamefont {M.}~\bibnamefont
  {Fr{\'\i}as-P{\'e}rez}}, \bibinfo {author} {\bibfnamefont {L.}~\bibnamefont
  {Tagliacozzo}},\ and\ \bibinfo {author} {\bibfnamefont {M.~C.}\ \bibnamefont
  {Ba{\~n}uls}},\ }\href@noop {} {\bibfield  {journal} {\bibinfo  {journal}
  {arXiv preprint arXiv:2308.04291}\ } (\bibinfo {year} {2023})}\BibitemShut
  {NoStop}%
\bibitem [{\citenamefont {Ba\~nuls}\ \emph {et~al.}(2009)\citenamefont
  {Ba\~nuls}, \citenamefont {Hastings}, \citenamefont {Verstraete},\ and\
  \citenamefont {Cirac}}]{PhysRevLett.102.240603}%
  \BibitemOpen
  \bibfield  {author} {\bibinfo {author} {\bibfnamefont {M.~C.}\ \bibnamefont
  {Ba\~nuls}}, \bibinfo {author} {\bibfnamefont {M.~B.}\ \bibnamefont
  {Hastings}}, \bibinfo {author} {\bibfnamefont {F.}~\bibnamefont
  {Verstraete}},\ and\ \bibinfo {author} {\bibfnamefont {J.~I.}\ \bibnamefont
  {Cirac}},\ }\href {https://doi.org/10.1103/PhysRevLett.102.240603} {\bibfield
   {journal} {\bibinfo  {journal} {Phys. Rev. Lett.}\ }\textbf {\bibinfo
  {volume} {102}},\ \bibinfo {pages} {240603} (\bibinfo {year}
  {2009})}\BibitemShut {NoStop}%
\bibitem [{\citenamefont {Müller-Hermes}\ \emph {et~al.}(2012)\citenamefont
  {Müller-Hermes}, \citenamefont {Cirac},\ and\ \citenamefont
  {Bañuls}}]{Muller-Hermes_2012}%
  \BibitemOpen
  \bibfield  {author} {\bibinfo {author} {\bibfnamefont {A.}~\bibnamefont
  {Müller-Hermes}}, \bibinfo {author} {\bibfnamefont {J.~I.}\ \bibnamefont
  {Cirac}},\ and\ \bibinfo {author} {\bibfnamefont {M.~C.}\ \bibnamefont
  {Bañuls}},\ }\href {https://doi.org/10.1088/1367-2630/14/7/075003}
  {\bibfield  {journal} {\bibinfo  {journal} {New Journal of Physics}\ }\textbf
  {\bibinfo {volume} {14}},\ \bibinfo {pages} {075003} (\bibinfo {year}
  {2012})}\BibitemShut {NoStop}%
\bibitem [{\citenamefont {Huang}\ \emph {et~al.}(2014)\citenamefont {Huang},
  \citenamefont {Chen}, \citenamefont {Kao},\ and\ \citenamefont
  {Xiang}}]{PhysRevB.89.201102}%
  \BibitemOpen
  \bibfield  {author} {\bibinfo {author} {\bibfnamefont {Y.-K.}\ \bibnamefont
  {Huang}}, \bibinfo {author} {\bibfnamefont {P.}~\bibnamefont {Chen}},
  \bibinfo {author} {\bibfnamefont {Y.-J.}\ \bibnamefont {Kao}},\ and\ \bibinfo
  {author} {\bibfnamefont {T.}~\bibnamefont {Xiang}},\ }\href
  {https://doi.org/10.1103/PhysRevB.89.201102} {\bibfield  {journal} {\bibinfo
  {journal} {Phys. Rev. B}\ }\textbf {\bibinfo {volume} {89}},\ \bibinfo
  {pages} {201102} (\bibinfo {year} {2014})}\BibitemShut {NoStop}%
\bibitem [{\citenamefont {Strathearn}\ \emph {et~al.}(2018)\citenamefont
  {Strathearn}, \citenamefont {Kirton}, \citenamefont {Kilda}, \citenamefont
  {Keeling},\ and\ \citenamefont {Lovett}}]{Strathearn2018}%
  \BibitemOpen
  \bibfield  {author} {\bibinfo {author} {\bibfnamefont {A.}~\bibnamefont
  {Strathearn}}, \bibinfo {author} {\bibfnamefont {P.}~\bibnamefont {Kirton}},
  \bibinfo {author} {\bibfnamefont {D.}~\bibnamefont {Kilda}}, \bibinfo
  {author} {\bibfnamefont {J.}~\bibnamefont {Keeling}},\ and\ \bibinfo {author}
  {\bibfnamefont {B.~W.}\ \bibnamefont {Lovett}},\ }\href
  {https://doi.org/10.1038/s41467-018-05617-3} {\bibfield  {journal} {\bibinfo
  {journal} {Nature Communications}\ }\textbf {\bibinfo {volume} {9}},\
  \bibinfo {pages} {3322} (\bibinfo {year} {2018})}\BibitemShut {NoStop}%
\bibitem [{\citenamefont {Fux}\ \emph {et~al.}(2021)\citenamefont {Fux},
  \citenamefont {Butler}, \citenamefont {Eastham}, \citenamefont {Lovett},\
  and\ \citenamefont {Keeling}}]{Fux21}%
  \BibitemOpen
  \bibfield  {author} {\bibinfo {author} {\bibfnamefont {G.~E.}\ \bibnamefont
  {Fux}}, \bibinfo {author} {\bibfnamefont {E.~P.}\ \bibnamefont {Butler}},
  \bibinfo {author} {\bibfnamefont {P.~R.}\ \bibnamefont {Eastham}}, \bibinfo
  {author} {\bibfnamefont {B.~W.}\ \bibnamefont {Lovett}},\ and\ \bibinfo
  {author} {\bibfnamefont {J.}~\bibnamefont {Keeling}},\ }\href
  {https://doi.org/10.1103/PhysRevLett.126.200401} {\bibfield  {journal}
  {\bibinfo  {journal} {Phys. Rev. Lett.}\ }\textbf {\bibinfo {volume} {126}},\
  \bibinfo {pages} {200401} (\bibinfo {year} {2021})}\BibitemShut {NoStop}%
\bibitem [{\citenamefont {Lerose}\ \emph {et~al.}(2021)\citenamefont {Lerose},
  \citenamefont {Sonner},\ and\ \citenamefont {Abanin}}]{Lerose21}%
  \BibitemOpen
  \bibfield  {author} {\bibinfo {author} {\bibfnamefont {A.}~\bibnamefont
  {Lerose}}, \bibinfo {author} {\bibfnamefont {M.}~\bibnamefont {Sonner}},\
  and\ \bibinfo {author} {\bibfnamefont {D.~A.}\ \bibnamefont {Abanin}},\
  }\href {https://doi.org/10.1103/PhysRevX.11.021040} {\bibfield  {journal}
  {\bibinfo  {journal} {Phys. Rev. X}\ }\textbf {\bibinfo {volume} {11}},\
  \bibinfo {pages} {021040} (\bibinfo {year} {2021})}\BibitemShut {NoStop}%
\bibitem [{\citenamefont {Thoenniss}\ \emph
  {et~al.}(2023{\natexlab{a}})\citenamefont {Thoenniss}, \citenamefont
  {Lerose},\ and\ \citenamefont {Abanin}}]{Thoenniss_Nonequlibrium}%
  \BibitemOpen
  \bibfield  {author} {\bibinfo {author} {\bibfnamefont {J.}~\bibnamefont
  {Thoenniss}}, \bibinfo {author} {\bibfnamefont {A.}~\bibnamefont {Lerose}},\
  and\ \bibinfo {author} {\bibfnamefont {D.~A.}\ \bibnamefont {Abanin}},\
  }\href {https://doi.org/10.1103/PhysRevB.107.195101} {\bibfield  {journal}
  {\bibinfo  {journal} {Phys. Rev. B}\ }\textbf {\bibinfo {volume} {107}},\
  \bibinfo {pages} {195101} (\bibinfo {year} {2023}{\natexlab{a}})}\BibitemShut
  {NoStop}%
\bibitem [{\citenamefont {Thoenniss}\ \emph
  {et~al.}(2023{\natexlab{b}})\citenamefont {Thoenniss}, \citenamefont
  {Sonner}, \citenamefont {Lerose},\ and\ \citenamefont
  {Abanin}}]{Thoenniss_Efficient}%
  \BibitemOpen
  \bibfield  {author} {\bibinfo {author} {\bibfnamefont {J.}~\bibnamefont
  {Thoenniss}}, \bibinfo {author} {\bibfnamefont {M.}~\bibnamefont {Sonner}},
  \bibinfo {author} {\bibfnamefont {A.}~\bibnamefont {Lerose}},\ and\ \bibinfo
  {author} {\bibfnamefont {D.~A.}\ \bibnamefont {Abanin}},\ }\href
  {https://doi.org/10.1103/PhysRevB.107.L201115} {\bibfield  {journal}
  {\bibinfo  {journal} {Phys. Rev. B}\ }\textbf {\bibinfo {volume} {107}},\
  \bibinfo {pages} {L201115} (\bibinfo {year}
  {2023}{\natexlab{b}})}\BibitemShut {NoStop}%
\bibitem [{\citenamefont {Ng}\ \emph {et~al.}(2023)\citenamefont {Ng},
  \citenamefont {Park}, \citenamefont {Millis}, \citenamefont {Chan},\ and\
  \citenamefont {Reichman}}]{Ng23}%
  \BibitemOpen
  \bibfield  {author} {\bibinfo {author} {\bibfnamefont {N.}~\bibnamefont
  {Ng}}, \bibinfo {author} {\bibfnamefont {G.}~\bibnamefont {Park}}, \bibinfo
  {author} {\bibfnamefont {A.~J.}\ \bibnamefont {Millis}}, \bibinfo {author}
  {\bibfnamefont {G.~K.-L.}\ \bibnamefont {Chan}},\ and\ \bibinfo {author}
  {\bibfnamefont {D.~R.}\ \bibnamefont {Reichman}},\ }\href
  {https://doi.org/10.1103/PhysRevB.107.125103} {\bibfield  {journal} {\bibinfo
   {journal} {Phys. Rev. B}\ }\textbf {\bibinfo {volume} {107}},\ \bibinfo
  {pages} {125103} (\bibinfo {year} {2023})}\BibitemShut {NoStop}%
\bibitem [{\citenamefont {Kloss}\ \emph {et~al.}(2023)\citenamefont {Kloss},
  \citenamefont {Thoenniss}, \citenamefont {Sonner}, \citenamefont {Lerose},
  \citenamefont {Fishman}, \citenamefont {Stoudenmire}, \citenamefont
  {Parcollet}, \citenamefont {Georges},\ and\ \citenamefont
  {Abanin}}]{Kloss23}%
  \BibitemOpen
  \bibfield  {author} {\bibinfo {author} {\bibfnamefont {B.}~\bibnamefont
  {Kloss}}, \bibinfo {author} {\bibfnamefont {J.}~\bibnamefont {Thoenniss}},
  \bibinfo {author} {\bibfnamefont {M.}~\bibnamefont {Sonner}}, \bibinfo
  {author} {\bibfnamefont {A.}~\bibnamefont {Lerose}}, \bibinfo {author}
  {\bibfnamefont {M.~T.}\ \bibnamefont {Fishman}}, \bibinfo {author}
  {\bibfnamefont {E.~M.}\ \bibnamefont {Stoudenmire}}, \bibinfo {author}
  {\bibfnamefont {O.}~\bibnamefont {Parcollet}}, \bibinfo {author}
  {\bibfnamefont {A.}~\bibnamefont {Georges}},\ and\ \bibinfo {author}
  {\bibfnamefont {D.~A.}\ \bibnamefont {Abanin}},\ }\href
  {https://doi.org/10.1103/PhysRevB.108.205110} {\bibfield  {journal} {\bibinfo
   {journal} {Phys. Rev. B}\ }\textbf {\bibinfo {volume} {108}},\ \bibinfo
  {pages} {205110} (\bibinfo {year} {2023})}\BibitemShut {NoStop}%
\bibitem [{\citenamefont {N\'u\~nez Fern\'andez}\ \emph
  {et~al.}(2022)\citenamefont {N\'u\~nez Fern\'andez}, \citenamefont {Jeannin},
  \citenamefont {Dumitrescu}, \citenamefont {Kloss}, \citenamefont {Kaye},
  \citenamefont {Parcollet},\ and\ \citenamefont
  {Waintal}}]{PhysRevX.12.041018}%
  \BibitemOpen
  \bibfield  {author} {\bibinfo {author} {\bibfnamefont {Y.}~\bibnamefont
  {N\'u\~nez Fern\'andez}}, \bibinfo {author} {\bibfnamefont {M.}~\bibnamefont
  {Jeannin}}, \bibinfo {author} {\bibfnamefont {P.~T.}\ \bibnamefont
  {Dumitrescu}}, \bibinfo {author} {\bibfnamefont {T.}~\bibnamefont {Kloss}},
  \bibinfo {author} {\bibfnamefont {J.}~\bibnamefont {Kaye}}, \bibinfo {author}
  {\bibfnamefont {O.}~\bibnamefont {Parcollet}},\ and\ \bibinfo {author}
  {\bibfnamefont {X.}~\bibnamefont {Waintal}},\ }\href
  {https://doi.org/10.1103/PhysRevX.12.041018} {\bibfield  {journal} {\bibinfo
  {journal} {Phys. Rev. X}\ }\textbf {\bibinfo {volume} {12}},\ \bibinfo
  {pages} {041018} (\bibinfo {year} {2022})}\BibitemShut {NoStop}%
\bibitem [{\citenamefont {Erpenbeck}\ \emph {et~al.}(2023)\citenamefont
  {Erpenbeck}, \citenamefont {Lin}, \citenamefont {Blommel}, \citenamefont
  {Zhang}, \citenamefont {Iskakov}, \citenamefont {Bernheimer}, \citenamefont
  {N{\'{u} }{\~{n}}ez-Fern{\'{a}}ndez}, \citenamefont {Cohen}, \citenamefont
  {Parcollet}, \citenamefont {Waintal},\ and\ \citenamefont
  {Gull}}]{Erpenbeck_2023}%
  \BibitemOpen
  \bibfield  {author} {\bibinfo {author} {\bibfnamefont {A.}~\bibnamefont
  {Erpenbeck}}, \bibinfo {author} {\bibfnamefont {W.-T.}\ \bibnamefont {Lin}},
  \bibinfo {author} {\bibfnamefont {T.}~\bibnamefont {Blommel}}, \bibinfo
  {author} {\bibfnamefont {L.}~\bibnamefont {Zhang}}, \bibinfo {author}
  {\bibfnamefont {S.}~\bibnamefont {Iskakov}}, \bibinfo {author} {\bibfnamefont
  {L.}~\bibnamefont {Bernheimer}}, \bibinfo {author} {\bibfnamefont
  {Y.}~\bibnamefont {N{\'{u} }{\~{n}}ez-Fern{\'{a}}ndez}}, \bibinfo {author}
  {\bibfnamefont {G.}~\bibnamefont {Cohen}}, \bibinfo {author} {\bibfnamefont
  {O.}~\bibnamefont {Parcollet}}, \bibinfo {author} {\bibfnamefont
  {X.}~\bibnamefont {Waintal}},\ and\ \bibinfo {author} {\bibfnamefont
  {E.}~\bibnamefont {Gull}},\ }\bibfield  {journal} {\bibinfo  {journal}
  {Physical Review B}\ }\textbf {\bibinfo {volume} {107}},\ \href
  {https://doi.org/10.1103/physrevb.107.245135} {10.1103/physrevb.107.245135}
  (\bibinfo {year} {2023})\BibitemShut {NoStop}%
\bibitem [{\citenamefont {Rakovszky}\ \emph {et~al.}(2022)\citenamefont
  {Rakovszky}, \citenamefont {von Keyserlingk},\ and\ \citenamefont
  {Pollmann}}]{Rakovszky22}%
  \BibitemOpen
  \bibfield  {author} {\bibinfo {author} {\bibfnamefont {T.}~\bibnamefont
  {Rakovszky}}, \bibinfo {author} {\bibfnamefont {C.~W.}\ \bibnamefont {von
  Keyserlingk}},\ and\ \bibinfo {author} {\bibfnamefont {F.}~\bibnamefont
  {Pollmann}},\ }\href {https://doi.org/10.1103/PhysRevB.105.075131} {\bibfield
   {journal} {\bibinfo  {journal} {Phys. Rev. B}\ }\textbf {\bibinfo {volume}
  {105}},\ \bibinfo {pages} {075131} (\bibinfo {year} {2022})}\BibitemShut
  {NoStop}%
\bibitem [{\citenamefont {von Keyserlingk}\ \emph {et~al.}(2022)\citenamefont
  {von Keyserlingk}, \citenamefont {Pollmann},\ and\ \citenamefont
  {Rakovszky}}]{Keyserlingk21}%
  \BibitemOpen
  \bibfield  {author} {\bibinfo {author} {\bibfnamefont {C.}~\bibnamefont {von
  Keyserlingk}}, \bibinfo {author} {\bibfnamefont {F.}~\bibnamefont
  {Pollmann}},\ and\ \bibinfo {author} {\bibfnamefont {T.}~\bibnamefont
  {Rakovszky}},\ }\href {https://doi.org/10.1103/PhysRevB.105.245101}
  {\bibfield  {journal} {\bibinfo  {journal} {Phys. Rev. B}\ }\textbf {\bibinfo
  {volume} {105}},\ \bibinfo {pages} {245101} (\bibinfo {year}
  {2022})}\BibitemShut {NoStop}%
\bibitem [{\citenamefont {Azad}\ \emph {et~al.}(2023)\citenamefont {Azad},
  \citenamefont {Hallam}, \citenamefont {Morley},\ and\ \citenamefont
  {Green}}]{Azad}%
  \BibitemOpen
  \bibfield  {author} {\bibinfo {author} {\bibfnamefont {F.}~\bibnamefont
  {Azad}}, \bibinfo {author} {\bibfnamefont {A.}~\bibnamefont {Hallam}},
  \bibinfo {author} {\bibfnamefont {J.}~\bibnamefont {Morley}},\ and\ \bibinfo
  {author} {\bibfnamefont {A.~G.}\ \bibnamefont {Green}},\ }\href
  {https://doi.org/10.1038/s41598-023-35336-9} {\bibfield  {journal} {\bibinfo
  {journal} {Scientific Reports}\ }\textbf {\bibinfo {volume} {13}} (\bibinfo
  {year} {2023})}\BibitemShut {NoStop}%
\end{thebibliography}%

\end{document}